\documentclass[reprint,12pt]{elsarticle}
\usepackage{amsmath}
\usepackage[numbers]{natbib}
\journal{Chaos, Solitons \& Fractals}
\newcommand{\sech}{\mbox{sech}}
\newcommand{\Del}{\Delta}
\newcommand{\del}{\delta}
\newcommand{\al}{\alpha}
\newcommand{\ba}{\beta}

\newcommand{\vphi}{\varphi}

\newcommand{\bea}{\begin{eqnarray}}
\newcommand{\eea}{\end{eqnarray}}
\newcommand{\bes}{\begin{subequations}}
\newcommand{\ees}{\end{subequations}}

\usepackage{xcolor}

\begin{document}

\begin{frontmatter}

\title{Soliton molecules in Fermi-Pasta-Ulam-Tsingou lattice: Gardner equation approach}

\author[ku]{M. Kirane\corref{auth1}}
\cortext[auth1]{Corresponding author}
\ead{mokhtar.kirane@ku.ac.ae}
\author[ku,bdu]{S. Stalin}
\ead{stalin.seenimuthu@ku.ac.ae;stalin.cnld@gmail.com}
\author[bdu]{R. Arun}
\author[bdu]{M. Lakshmanan}
\ead{lakshman@cnld.bdu.ac.in}
\address[ku]{Department of Mathematics, College of Art and Sciences, Khalifa University of Science and Technology, Abu Dhabi, 127788, United Arab Emirates}
\address[bdu]{Department of Nonlinear Dynamics, School of Physics, Bharathidasan University, Tiruchirappalli - 620024, Tamilnadu, India}




\begin{abstract}
We revisit the Fermi-Pasta-Ulam-Tsingou lattice (FPUT) with quadratic and cubic nonlinear interactions in the continuous limit by deducing the Gardner equation. Through the Hirota bilinear method, the fundamental as well as multi-soliton, particularly two-, three- and four-soliton, solutions are obtained for the Gardner equation. Based on these multi-soliton solutions, we show the excitation of an interesting class of table-top soliton molecules in the FPUT lattice   through the velocity resonance mechanism.  Depending on the condition on the free parameters, which determine the bond length between the soliton atoms, we classify them as dissociated and synthetic type molecules. The main feature of the table-top soliton molecules is that they do not exhibit oscillations in the coalescence region. This property ensures that they are distinct from the soliton molecules, having retrieval force, of the nonlinear Schr\"odinger family of systems. Further, to study the stability of the soliton molecule we allow it to interact with a single (or multi) soliton(s). The asymptotic analysis shows that their structures remain constant, though the bond length varies, throughout the collision process. Then, from the relative separation distance calculation we also observe that the bond length of each of the soliton molecules is not maintained during soliton molecule-molecule interaction while their structures are preserved.  In addition,  we consider the FPUT lattice with quadratic nonlinear interaction and FPUT lattice with cubic nonlinearity as sub-cases and point out the nature of the soliton molecules for these cases also systematically. We achieve this based on the interconnections between the solutions of the Gardner, modified Korteweg-de Vries and Korteweg-de Vries equations. Finally, in order to validate the existence of all the soliton structures, especially reported for the Gardner equation in the continuous limit, we simulate the corresponding FPUT chain  and verified these various structures numerically and analytically as well. We believe that the present study can be extended to other integrable and non-integrable systems with applications in fluid dynamics, Bose-Einstein condensates, nonlinear optics, and plasma physics.          
\end{abstract}


\end{frontmatter}


\section{Introduction}
Soliton molecule (SM) is a bound soliton state which is formed at stable equilibrium distance where the interaction force is zero among the two solitons \cite{Rohrmann,fedor-epjplus,sano,fm1,grelu-ph1}. An important feature of SM is that the elementary solitons should have equal group velocities. 
Recently, the study on SMs in different areas of physics has attracted attention significantly due to their unique behaviours and are of immense interest for application in optical telecommunications \cite{Rohrmann,fedor-epjplus}.
For instance, it has been proposed that the SM in dispersion managed fiber system is useful for enhancing the information carrying capacity of the fiber \cite{fm1,fm2} by considering this novel soliton state as next level of alphabet in the binary coding scheme \cite{Rohrmann,fedor-epjplus}. Such SMs have also been observed in Bose-Einstein condensates (BECs) \cite{khawaja1,santos1,santos2}, optical systems \cite{fm1,haus1,solli,liu,nithya,haus2,khawaja2}, and other closely related fields \cite{crasovan,fm4,khawaja3,matre,nithya1}.
In the literature, soliton/breather molecules or soliton molecular complexes have been investigated both in conservative \cite{na1,gelash,andrey,shin,lou1,lou2,lou3,lou4,lou5,bli,wang,zhang,zhou1} and dissipative systems \cite{segev,fm1,khawaja4,marconi,liang}. In the conventional conservative case they show fascinating  dynamical features which are very useful to understand their properties in the dissipative case, where they exhibit rich dynamical features. For instance, dissipative SMs or the so called photonic molecules can show dynamics similar to matter molecules such as synthesis and vibration \cite{nithya,nithya1}. Due to the existence of such fascinating properties, the soliton molecules have been investigated rigorously both theoretically and experimentally \cite{segev,fm1,khawaja4,marconi,liang}.

 Besides the above interesting works on SMs, it is important to point out that the study of them goes back to the earlier literature on the investigation of bound states of solitons \cite{ostrov,karpman,gord,na1,fm3,sm1,sm2,afanasjev1,afanasjev2,fl1,fl2,fl3}. The force between solitons was first studied in Ref. \cite{karpman} using perturbation analysis.
 It was shown that the force of interaction depends on the spatial separation and the phase difference between the solitons \cite{gord,fm3}. It was also shown there that the in-phase solitons attract each other whereas out-of-phase solitons repel \cite{fm3}. Then, using perturbation approach, the existence of bound soliton state of two solitons has been exhibited by Malomed in \cite{sm2}. However, it has been numerically demonstrated that such bound states are unstable \cite{afanasjev1,afanasjev2}. In Ref. \cite{soto}, Akhmediev et al. displayed the possibility of observing stable bound states. Later in Ref. \cite{khawaja1}, the formula for force of interaction between the solitons for arbitrary choice of soliton parameters was obtained by extending the Gordon's formula of force of interaction between the solitons \cite{gord}. From the above contemporary interest on soliton molecules we realized that one has to essentially understand the concept of SMs even at the fundamental level, where non-integrable/integrable models play a decisive role in governing the evolution of these systems. Further, it is also very important to investigate the formation mechanism of SMs, stability against perturbations and their dynamics with respect to various physical conditions. Based on the above studies, we conclude that to the best of our knowledge the problem of existence of soliton molecules in the continuous limit of the famous Fermi-Pasta-Ulam-Tsingou lattice, with quadratic-cubic nonlinear interactions, and their dynamical properties have not been addressed so far in the literature. To unveil this, we consider multi-soliton solutions of the completely integrable Gardner equation, which is deduced from the FPUT lattice in the asymptotic continuous limit. Then, we validate our results, especially for the Gardner equation case, by numerically simulating the corresponding FPUT chain.   
 
To address our main objective, we consider the widely celebrated FPUT lattice model with quadratic and cubic nonlinear nearest neighbor interactions. The corresponding equation of motion of the FPUT lattice is given by \cite{td,lakshmanan}
\bea
m\frac{d^2y_i}{dt^2}=f(y_{i+1}-y_i)-f(y_i-y_{i-1}), ~~i=1,2,....,N. \label{fput}
\eea
In the above, $y_i$ is the displacement of the $i$-th mass point from the equilibrium position,  and $m$ is the mass of a point particle. The lattice model (\ref{fput}) was proposed to verify the ergodicity hypothesis of statistical mechanics \cite{td,lakshmanan,berman} in which it was expected that any weak nonlinear interaction between particles obey equipartition of energy. On the contrary, the result was opposite: the energy was periodically returning to the initially excited mode. This famous FPUT paradox was later resolved \cite{td,berman} and it leads to new concept known as ``integrability" of nonlinear partial differential equations. The following forms of forces were considered in Ref. \cite{berman}, besides broken linear type, to check the fundamental laws of statistical mechanics, 
\bes\begin{equation}
f(y)=\alpha y^2+ky, \end{equation}
and\begin{equation}
f(y)=ky+\beta y^3.
\end{equation}\ees 
In the above, the constants $\alpha$ and $\beta$ are the strengths of quadratic and cubic nonlinear interactions between the mass point particles, respectively. If the nonlinearities are absent $\alpha=0$, and $\beta=0$, then $f(y)=ky$, where $k$ is the force constant. Then the equation of motion (\ref{fput}) becomes
\bea
m\frac{d^2y_i}{dt^2}=k(y_{i+1}-2y_i+y_{i-1}).
\eea 
As it is well known from the earlier studies \cite{lakshmanan}, by applying a normal mode decomposition \cite{berman}, one can easily understand that no energy sharing will take place
among the constituent modes and they oscillate independently in a simple harmonic manner. On the other hand, when $\alpha \neq 0$, and $\beta=0$, for example, one can expect the phenomenon of recurrence of energy to the initial modes \cite{lakshmanan}. It was explained by the concept of soliton by solving the Korteweg-de Vries (KdV) equation, which was deduced from Eq. (\ref{fput}) in the continuous limit, numerically \cite{kruskal} with appropriate initial and boundary conditions. We note that  for $\alpha=0$, and $\beta\neq 0$, that is FPUT model (\ref{fput}) with cubic nonlinearity, the modified KdV (mKdV) equation is deduced in the continuous limit.   

In the present study, we investigate the model (\ref{fput}), with $\alpha \neq 0$, and $\beta\neq 0$, in the continuous limit. In this approximation limit, we deduce the following Gardner equation,
\begin{eqnarray}
	4v_t+6v^2v_x+12\mu vv_x-v_{xxx}=0, ~~~\mu=-\sqrt{\frac{2a_1}{3}}<0,  \label{1}
\end{eqnarray}
and derive multi-soliton solutions for this extended KdV equation. Then, through soliton solutions of the Gardner equation, we show the excitation of an interesting class of soliton molecules in the FPUT lattice (\ref{fput}). These SMs are denoted as bi-, tri-, and quad-types with respect to the number of constituent soliton atoms. Then, based on the calculation of bond lengths, we classify them as dissociated and synthetic type molecules. Further, we analyze the stability and collision properties of these soliton molecules  by considering three different situations: 
(i) collision between a bi-SM and a fundamental soliton, (ii) interaction between a bi-SM and two individual solitons, and (iii) collision between two bi-SMs. The asymptotic analysis reveals that the molecules maintain their structures  against collisions and thereby confirming the elastic nature of collision. Besides these, we also indicate the existence of SMs in the other sub-cases, namely the modified KdV and KdV equations, which can be deduced from the respective continuous limit of model (\ref{fput}) with cubic interaction and model (\ref{fput}) with quadratic nonlinear interaction. The soliton solutions for these fundamental integrable equations can be deduced by knowing the soliton solutions of the Gardner equation. We remark that such explicit forms of soliton solutions, derived by this way for those fundamental integrable models, are new to the literature. 

The rest of the paper is organized as follows. In the next section, 
we derive multi-soliton solutions of the Gardner equation through the Hirota bilinear method and demonstrate the standard elastic collision among them through appropriate asymptotic analysis. The soliton molecules of the Gardner equation and their associated properties are explained in detail in section 3. Section 4 deals with the collision and stability properties of SMs. These are studied by performing an appropriate asymptotic analysis. Sections 5 and 6 mainly report the existence of SMs in the modified KdV and KdV equations, respectively. In Section 7, we numerically simulate the FPUT chain corresponding to the Gardner equation and show the existence of various soliton structures which are admitted by the Gardner equation. We summarize the results in Section 8. Further, in Appendix A, we recall how to deduce the Gardner equation from the model (\ref{fput}) in the asymptotic continuous limit $a\rightarrow 0$, where $a$ is a lattice parameter. Also, we show there how the solutions of the mKdV and KdV equations can be derived from the solution of the Gardner equation. In Appendix B, we present the three-soliton solution and an asymptotic analysis to ensure the occurrence of elastic collision among the three solitons.          
\section{Soliton solutions of the Gardner equation}
To show that the FPUT lattice (\ref{fput}) in the continuous limit can admit specific table-top soliton excitation, we intend to derive the soliton solutions for the Gardner Eq. (\ref{1}) by bilinearizing it through the following dependent variable transformation 
\begin{eqnarray}
	v(x,t)=\frac{G}{F},~ ~G\equiv G(x,t),~F\equiv F(x,t).\label{2}
\end{eqnarray}
 In the above, both the unknown functions $G$ and $F$ are real functions. By substituting the transformation (\ref{2}) in Eq. (\ref{1}) we obtain the bilinear forms of it and they turn out to be
\begin{eqnarray}
	(4D_t-D_x^3)G\cdot F=0, ~ D_x^2 F\cdot F=-2G^2-4\mu G\cdot F. \label{15}
\end{eqnarray}
Here, $D_x$ and $D_t$ are the standard bilinear operators \cite{hirota}. 
To find the unknown functions $G$ and $F$ one has to solve a system of linear partial differential equations which arise while considering the standard series expansions,
\begin{eqnarray}
	G=b+\epsilon g_1+\epsilon^2g_2+\epsilon^3 g_3+...,~
	F=1+\epsilon f_1+\epsilon^2 f_2+\epsilon^3 f_3+\epsilon^4 f_4+...,
	\label{4}
\end{eqnarray}
in the bilinear equations (\ref{15}). In the above, $b$ is a constant which can be determined later during the solution construction process. Hereafter, one has to follow the standard Hirota's procedure to obtain the multi-soliton solutions of the Gardner equation (\ref{1}).  
\subsection{One-soliton solution: Excitation of table-top soliton in FPUT lattice (\ref{fput})}
We derive the fundamental soliton solution of the Gardner equation (\ref{1}) by following the standard procedure described above. The form of one-soliton solution is obtained as
\begin{eqnarray}
	v(x,t)=\frac{\epsilon g_1 }{1+\epsilon f_1+\epsilon^2 f_2}=\frac{e^{\eta_1}}{1+e^{\eta_1+\delta_1}+e^{2\eta_1+\Delta_1}}, \label{17}
\end{eqnarray}
where  $\eta_1=k_1x+\frac{k_1^3}{4}t+\eta_1^{(0)}$,
$\displaystyle{e^{\del_1}= -\frac{2\mu}{k_1^2}}$, $\displaystyle{e^{\Del_1}=\frac{4\mu^2-k_1^2}{4k_1^4}}$, $\eta_1^{(0)}=\log\epsilon$, and $b=0$, which is true for the higher-order soliton solutions too. Here, the wave number $k_1$ and the phase $\eta_1^{(0)}$ are real constants. 

\begin{figure}[]	
	\centering
	\includegraphics[width=0.5\linewidth]{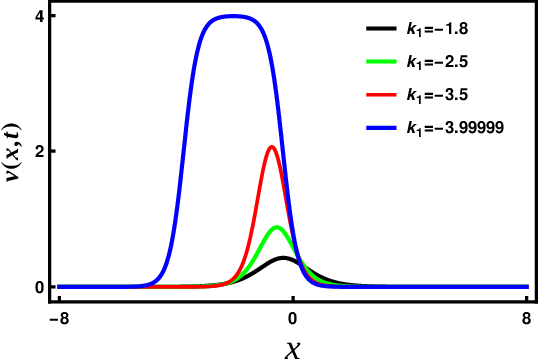}
	\caption{Various profiles of the fundamental soliton of  the Gardner equation (\ref{fput}). A large amplitude table-top soliton excitation (blue line) is shown for  $k_1=-3.99999$. }
	\label{fg1}
\end{figure}

To bring out the properties associated with the solution (\ref{17}), we rewrite it in a hyperbolic form. This action gives 
\bea
v(x,t)=\frac{k_1^2}{-2\mu+\sqrt{4\mu^2-k_1^2}\cosh(\eta_1+\frac{\Del_1}{2})}. \label{18}
\eea
From the above, one can easily find that the two real parameters $\mu$ (which is always negative, see Eq. (\ref{1})) and $k_1$ predominantly influence  the structure of the Gardner soliton apart from a constant parameter $\eta_1^{(0)}$.    
The amplitude and velocity of the fundamental soliton are identified from the expression (\ref{18}) as $-k_1^2/(2\mu)$, and $c=k_1^2/4$, respectively. The central  position of the soliton is also deduced from the same expression as
$\frac{\eta_1^{(0)}}{k_1}+\frac{\Delta_1}{2k_1}=\frac{\eta_1^{(0)}}{k_1}+\frac{1}{2k_1}\log\frac{4\mu^2-k_1^2}{4k_1^4}$. The solution (\ref{18}) admits a non-singular, real and smooth solution for $4\mu^2>k_1^2$ (Otherwise, one gets a complex solution which is in general not acceptable). The latter restriction limits both amplitude and velocity of the Gardner soliton. However, a special feature of this restriction is that it makes the fundamental soliton solution (\ref{18}) to admit a special flat-top profile, as it is drawn in Fig. \ref{fg1} by a blue line for a specific parameter value of $k_1$. In this limit, the fundamental soliton possesses maximum amplitude and velocity. To get such a limiting large amplitude table-top soliton we fix $k_1=-3.99999$ and the other parameter values are fixed as $\eta_1^{(0)}=0$ and $a_1=6$. We note that one can also choose positive numerical value for $k_1$.  If so, then the soliton localization will be along the $+x$ axis. This flat-top structure, arises in the Gardner Eq. (\ref{1}), can be visualized as the combination of a kink and an anti-kink solitons as it was shown in \cite{slyunyaev1,chow,grimshaw1}, where the analytical form of the table-top fundamental soliton was rewritten as the combination of analytical forms of kink-anti kink solitons. This confirms that the simultaneous appearance of competing quadratic and cubic nonlinear interactions in the FPUT lattice induces a special kind of flat-top soliton excitation in the asymptotic continuum limit as well as in the discrete limit. The existence of such a flat-top soliton excitation is confirmed in the FPUT lattice, in Section 7 numerically. We believe that this possibility has not been discussed earlier in the FPUT lattice, except in Ref. \cite{kudryashov}, where the authors have considered a particular form of solitary wave solution for their numerical investigation. Besides this, we also observe that the solution (\ref{18})  displays the usual KdV like sech soliton profiles for lower values of $k_1$. In Fig. \ref{fg1}, the transition from sech-like profile to flat-top soliton profile is clearly demonstrated by considering the various values of $k_1$.  

In addition to the above, another interesting feature associated with the solution (\ref{18}) is that it has the amplitude-dependent velocity property. To visualize this, the expression (\ref{18}) is rewritten as follows:
\bea
v(x,t)=\frac{2c}{-\mu+\sqrt{\mu^2-c}\cosh\big[2\sqrt{c}(x-ct+\frac{\Del_1}{4\sqrt{c}}+\frac{+\eta_1^{(0)}}{2\sqrt{c}})\big]}. 
\eea
From the above expression, one can easily confirm that the velocity explicitly appears in the amplitude part of the soliton. Therefore, the taller Gardner soliton will travel faster than the smaller one, as this property was identified a long time ago in the case of KdV equation \cite{lakshmanan,kruskal}.  
It is important to note that a large amplitude soliton in table-top form has been extensively discussed in the context of internal solitons in fluid dynamics \cite{grimshaw1,ostro,grimshaw2}, where several classical integrable  models and their variants have been considered to study the internal solitons with different environment conditions. In addition to this, soliton and breather solutions have also been reported for another form of the extended KdV/Gardner equation \cite{chow,slyunyaev2}. In the present case, we consider the coefficients of quadratic nonlinearity and dispersion as negative in order to distinguish our work from the others \cite{slyunyaev1,chow,grimshaw1}. However, one can get the soliton solutions of the Gardner equation (\ref{1}) with positive dispersion by scaling $x\rightarrow -x$.      
\subsection{Two-soliton solution}
\begin{figure}[]
	\centering
	\includegraphics[width=0.72\linewidth]{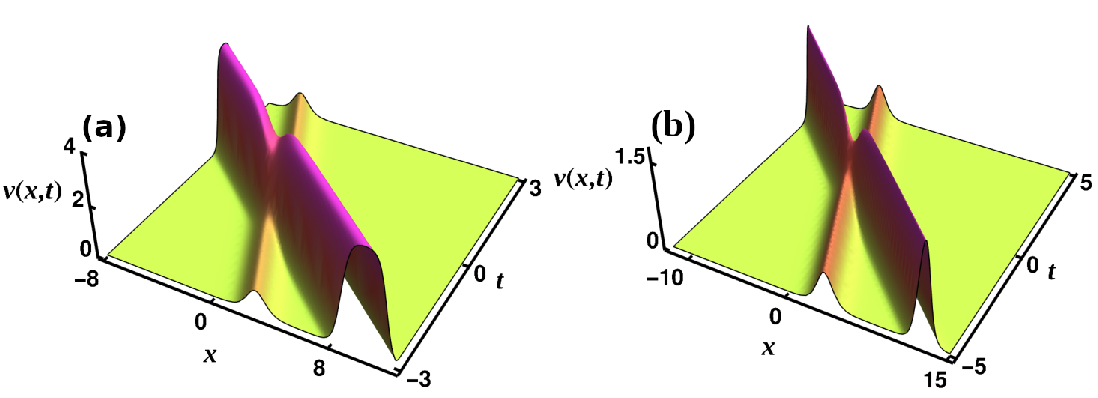}\\
	\includegraphics[width=0.36\linewidth]{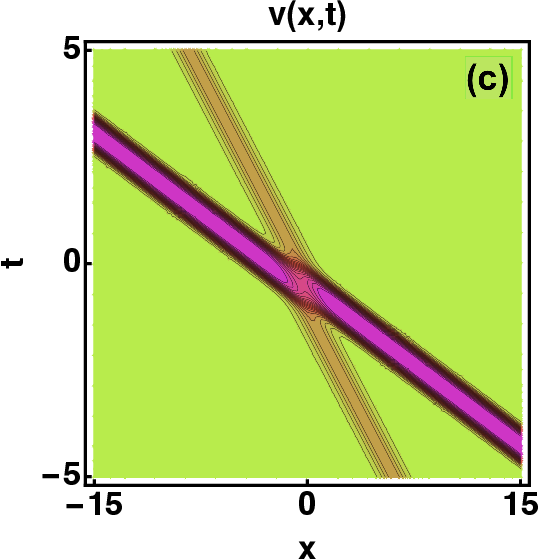}~~\includegraphics[width=0.36\linewidth]{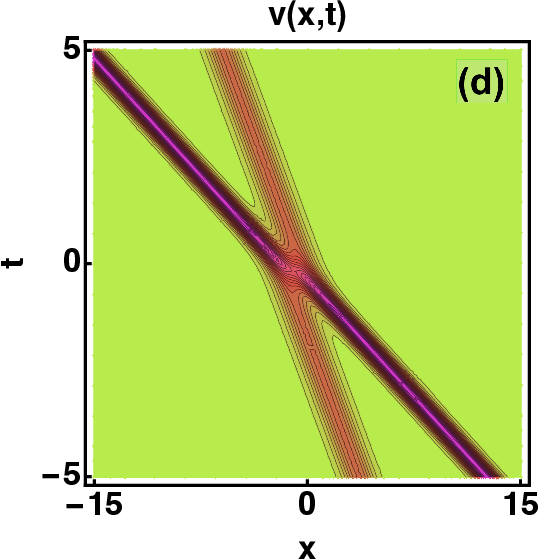}
	\caption{The figure (a) shows that an elastic collision between a table-top soliton and a sech-type soliton. The figure (b) illustrates the collision among the two sech-type solitons. Their corresponding space-time plots are given in figures (c), and (d), respectively, ensuring the phase shifts of solitons.   }
	\label{fg2}
\end{figure}

\begin{figure}[]
	\centering
	\includegraphics[width=0.8\linewidth]{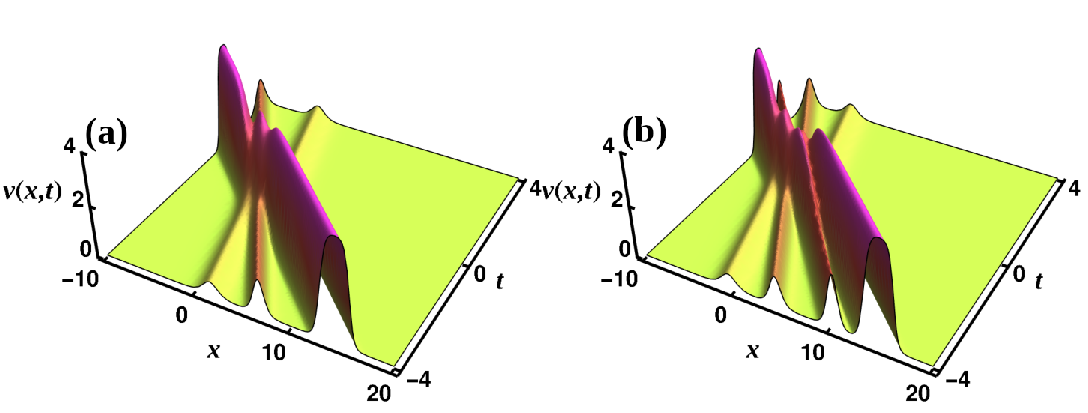}
	\caption{The figure (a) demonstrates an elastic collision between the three Gardner solitons. To bring out this collision we fix the parameter values as  $k_1=-2$, $k_2=-3$, $k_3=-3.9999$, $\eta_1^{(0)}=1.5$, $\eta_2^{(0)}=0$, and $\eta_3^{(0)}=-1$. The figure (b) illustrates an elastic collision between the four Gardner solitons. To draw this figure we fix the parameter values as $k_1=-2$, $k_2=-3$, $k_2=-3.5$, $k_4=-3.9999$, $\eta_1^{(0)}=0$, $\eta_2^{(0)}=0$, $\eta_3^{(0)}=1$, and $\eta_4^{(0)}=2$.}
	\label{fg3}
\end{figure}

To get the two-soliton solution of Eq. (\ref{1}), one has to find the explicit forms of the unknown functions in the truncated series, $G=\epsilon g_1+\epsilon^2 g_2+\epsilon^3 g_3$ and $f=1+\epsilon f_1+\epsilon^2 f_2+\epsilon^3 f_3+\epsilon^4 f_4$. By doing so, the resultant forms constitute the two-soliton solution for Eq. (\ref{1}) and it reads 
\bes
\begin{eqnarray}
	&&\hspace{-0.5cm}v=\frac{1}{F}\bigg(e^{\eta_1}+e^{\eta_2}+e^{\eta_1+\eta_2+\chi_1}+e^{2\eta_1+\eta_2+\chi_{11}}+e^{\eta_1+2\eta_2+\chi_{12}}\bigg),\label{20a}\\
	&&\hspace{-0.5cm}F=1+e^{\eta_1+\delta_1}+e^{\eta_2+\delta_{2}}+e^{2\eta_1+\Delta_1}+e^{2\eta_2+\Delta_{2}}+e^{\eta_2+\eta_1+\delta_{11}}\nonumber\\
	&&~+e^{2\eta_1+\eta_2+\delta_{21}}+e^{\eta_1+2\eta_2+\delta_{22}}+e^{2\eta_1+2\eta_2+\delta_{31}}, \label{20b}
\end{eqnarray}\ees
where $\eta_j=k_jx+\frac{k_j^3}{4}t+\eta_j^{(0)}$, $j=1,2,$

\bea
&&\hspace{-0.5cm}e^{\del_j}= -\frac{2\mu}{k_j^2}, ~ e^{\Del_j}=\frac{4\mu^2-k_j^2}{4k_j^4},~e^{\delta_{11}}=\frac{2\big[2\mu^2(k_1^2+k_2^2)-k_1^2k_2^2\big]}{k_1^2k_2^2(k_1+k_2)^2},\nonumber\\
&&\hspace{-0.5cm} e^{\chi_1}=-\frac{2\mu(k_1-k_2)^2}{k_1^2k_2^2},~e^{\chi_{1j}}=\frac{(4\mu^2-k_j^2)(k_1-k_2)^2}{4k_j^4(k_1+k_2)^2},~j=1,2,\nonumber\\
&&\hspace{-0.5cm} e^{\delta_{21}}=-\frac{\mu(4\mu^2-k_1^2)(k_1-k_2)^2}{2k_1^4k_2^2(k_1+k_2)^2},~~e^{\delta_{22}}=-\frac{\mu(4\mu^2-k_2^2)(k_1-k_2)^2}{2k_1^2k_2^4(k_1+k_2)^2},\nonumber\\
&&\hspace{-0.5cm} \text{and}~~  e^{\delta_{31}}=\frac{(4\mu^2-k_1^2)(4\mu^2-k_2^2)(k_1-k_2)^4}{16k_1^4k_2^4(k_1+k_2)^4}.\nonumber
\eea
The structure and dynamics of the two-soliton solution are mainly described by the four real constants $k_j$, and $\eta_j^{(0)}$, $j=1,2$, apart from the system parameter $\mu$. A typical overtaking collision between a table-top soliton and a smooth sech-type soliton is displayed in Fig. \ref{fg2}(a) for the parameter values $k_1=-2.5$, $k_2=-3.9999$, $a_1=-6$, $\eta_1^{(0)}=0$ and $\eta_2^{(0)}=-1$. This figure shows that the table-top soliton, which has a larger speed, collides with a slowly moving sech-type soliton at $x=0$. Then it simply overtakes the smaller one and they reemerge without change in size, shape and velocity, except for a finite phase shift, thereby confirming the elastic nature of the collision. The final outcome of the collision is that each of the solitons suffers only a finite phase shift. The phase shifts corresponding to the two solitons are related as follows: $\Phi_2=\phi_2^+-\phi_2^-=\frac{1}{2}\log\frac{(4\mu^2-k_2^2)(k_1-k_2)^4}{4k_2^4(k_1+k_2)^4}-\frac{1}{2}\log\frac{(4\mu^2-k_2^2)}{4k_2^4}=\frac{1}{2}\log\frac{(k_1-k_2)^4}{(k_1+k_2)^4}=-\Phi_1=\phi_1^+-\phi_1^-=\frac{1}{2}\log\frac{(4\mu^2-k_1^2)}{4k_1^4}-\frac{1}{2}\log\frac{(4\mu^2-k_1^2)(k_1-k_2)^4}{4k_1^4(k_1+k_2)^4}$, where $\phi_j^+$ and $\phi_j^-$, $j=1,2$, are the respective phases of the two individual solitons before and after collision. We also depict an elastic collision among the two sech-type solitons in Fig. \ref{fg2}(b) with the parameter values $k_1=-2.1$, $k_2=-3.2999$, $\eta_1^{(0)}=0$ and $\eta_2^{(0)}=-1$. The space-time graphs corresponding to Figs. \ref{fg2}(a) and \ref{fg2}(b) are demonstrated in Figs. \ref{fg2}(c) and \ref{fg2}(d), respectively, to ensure the Gardner solitons always change their direction of propagation after collision.  These collision scenarios, from both Figs. \ref{fg2} (a) and (b), substantiate the famous observation of FPUT recurrence phenomenon \cite{kruskal} in the dynamics of an anharmonic lattice. This kind of collision property was also observed in a two-layer fluid with a density jump at the
interface \cite{slyunyaev1,ostro} where the two internal solitons interact elastically. We also observe such a shape preserving  collision even in the higher-order solitons cases as well. To ensure this we derive the higher-order soliton solutions, specifically the three and four soliton solutions, of the Gardner equation (\ref{1}). The three-soliton solution is explicitly given in the Appendix B while we have omitted the details of the four-soliton solution because of its complicated form. These high-order solutions will also help us later to investigate our primary objective on soliton molecules. An elastic collision among the three-solitons is demonstrated  in  Fig. \ref{fg3}(a). Similarly, in the four-soliton case also we observed an elastic collision among the solitons. It is demonstrated in Fig. \ref{fg3}(b). These figures clearly confirm that the Gardner solitons re-emerge from the collision scenario without being affected in their structures except for the phase shifts. One can also confirm this by doing an appropriate asymptotic analysis as we have done in the Appendix B for the case of the three-soliton solution.

\section{Soliton molecules}
As we have pointed out in the introductory section, the soliton molecule is a bound state of two or more elementary soliton atoms that are separated by a finite equilibrium distance.  In general, the bond length or the separation distance between two atoms is relatively larger than the width of a single soliton atom \cite{afanasjev1}. If the distance between the constituent atoms is too larger than a single soliton then the interaction among them is weak. In this case, the SM can be called as a dissociated molecule \cite{gelash}. On the other hand, depending on the force (attractive/repulsive) of interaction and for a smaller separation distance, a coalescence phenomenon of solitons will happen in the overlapping region, where the interaction among the solitons is strong leading to the formation of interference patterns. In this situation, the corresponding  molecular state can be designated as a synthetic molecule. It is very important to note that the constituents of a SM possess exactly equal or almost equal group velocities so that the entire molecular structure propagates like a single entity, in which the soliton atoms co-propagate in parallel. Such a velocity resonance condition is essential for the formation of a soliton molecule.           

In the present conservative case, we show the existence of table-top SMs in the continuous limit of the FPUT lattice (\ref{fput}) based on the multi-soliton solutions of the Gardner equation. In particular, we obtain the fundamental/bi-, tri-, and quad-SMs from the two-, three-, and four-soliton solutions, respectively. We characterize the SMs based on the following two points: (i) Bond length: Analysis of the spatial separation distance between the atoms. For instance, the spatial relative separation distance of the bi-soliton molecule can be obtained as $\Delta x_{12}=\phi_1-\phi_2$, where $\phi_1$ and $\phi_2$ are the phases of the solitons. (ii) Amplitude and width: During space-time translation, the molecular structure either varies or remains constant throughout the evolution.   To verify this, one has to trace out the changes in the amplitude and width of each of the solitons with respect to space and time.  

\begin{figure}[]
	\centering
	\includegraphics[width=0.9\linewidth]{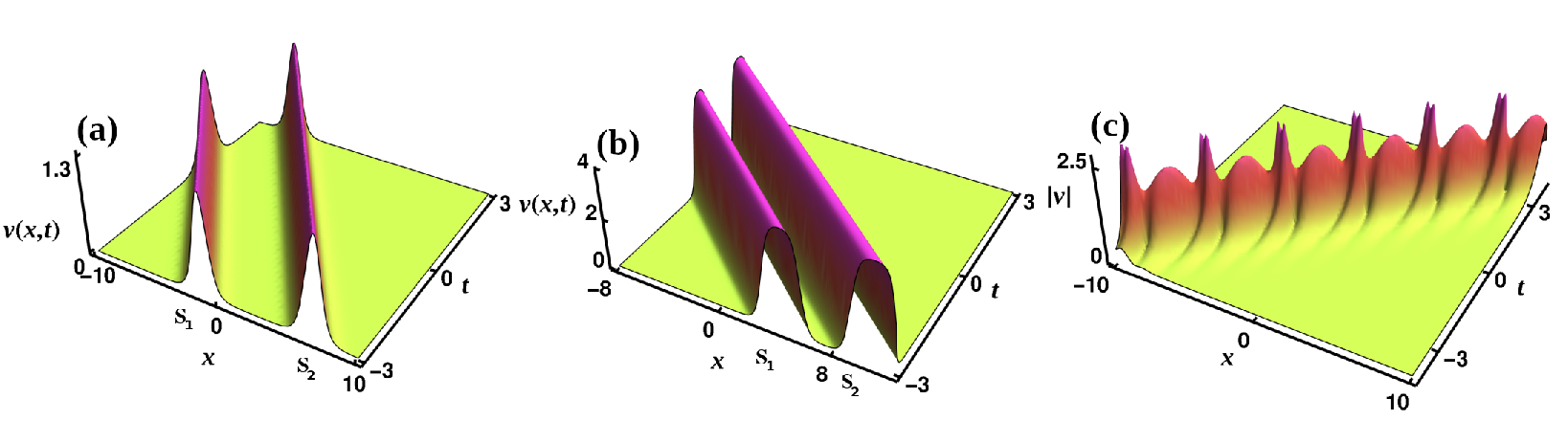}
	\caption{The figure (a) corresponds to a sech-type dissociated soliton molecule obeying the condition $\eta_1^{(0)}(=0)\neq \eta_2^{(0)}(=2)$. A synthetic table-top soliton molecule is displayed in Fig. (b) with the condition $\eta_1^{(0)}= \eta_2^{(0)}=0$. A special kind of breathing molecule state is illustrated in Fig. (c) by considering the wave number resonance condition $k_2=1+2i=k_1^*$ along with a complex phase constants  $\eta_2^{(0)}=\eta_1^{(0)*}=0$.}
	\label{fg4}
\end{figure}

To visualize the analytical form of the fundamental SM, we rewrite the two-soliton solution (\ref{20a})-(\ref{20b}) of the Gardner equation in hyperbolic form which turns out to be   
\bes\bea
v(x,t)&=&\frac{1}{D}\bigg(k_2^2(4\mu^2-k_1^2)^\frac{1}{2}\cosh(\eta_1+\phi_1)+k_1^2(4\mu^2-k_2^2)^\frac{1}{2}\cosh(\eta_2+\phi_2)\nonumber\\&&-2\mu(k_1^2-k_2^2)\bigg),\label{29a}\\
D&=&\frac{c_1}{2}(4\mu^2-k_1^2)^\frac{1}{2}(4\mu^2-k_2^2)^\frac{1}{2}\cosh(\eta_1+\eta_2+\phi_1+\phi_2)\nonumber\\&&+\frac{1}{2c_1}(4\mu^2-k_1^2)^\frac{1}{2}(4\mu^2-k_2^2)^\frac{1}{2}\cosh(\eta_1-\eta_2+\phi_1-\phi_2)\label{29b}\\
&&-2\mu(4\mu^2-k_1^2)^\frac{1}{2}\cosh(\eta_1+\phi_1)-2\mu(4\mu^2-k_2^2)^\frac{1}{2}\cosh(\eta_2+\phi_2)+c_2,\nonumber
\eea\ees
where $\eta_1=k_1(x+v_1t+\frac{\eta_1^{(0)}}{k_1})$, $\eta_2=k_2(x+v_2t+\frac{\eta_2^{(0)}}{k_2})$, $v_1=\frac{k_1^2}{4}$, \\$c_1=\frac{k_1-k_2}{k_1+k_2}$, $v_2=\frac{k_2^2}{4}$, $\phi_1=\frac{1}{2}\log\frac{(4\mu^2-k_1^2)(k_1-k_2)^2}{4k_1^4(k_1+k_2)^2}$, $\phi_2=\frac{1}{2}\log\frac{(4\mu^2-k_2^2)(k_1-k_2)^2}{4k_2^4(k_1+k_2)^2}$,
 $c_2=\frac{2[2\mu^2(k_1^2+k_2^2)-k_1^2k_2^2]}{k_1^2-k_2^2}$. The formation of bi-SM can happen if the two non-stationary solitons of the above two-soliton solution obey the velocity resonance condition \cite{gelash,lou1},
\begin{equation}
\frac{\omega_1}{\omega_2}=\frac{k_1^3}{k_2^3}~~ \text{or}~~\frac{v_1}{v_2}=\frac{k_1^2}{k_2^2}\rightarrow 1,~ \text{when} ~ k_2\rightarrow  k_1. \label{30}
\end{equation}
This implies that the two solitons of the same kind tends to align parallely and propagate like a single molecular structure, as it is displayed in Figs. \ref{fg4}(a) and \ref{fg4}(b), with the velocity $v_1\approx v_2=v_{\text{mol}}$. 

Further, the bond length between the individual solitons  of the bi-SM is deduced from the expression (\ref{29a})-(\ref{29b}) by calculating the relative separation distance between the atoms from soliton 2 to soliton 1. This results in the expression
\bea
\Delta x_{21} &=&\frac{\eta_2^{(0)}}{k_2}+\frac{\phi_2}{k_2}-
\frac{\eta_1^{(0)}}{k_1}-\frac{\phi_1}{k_1}=\frac{1}{2k_2}\log\frac{(4\mu^2-k_2^2)(k_1-k_2)^2}{4k_2^4(k_1+k_2)^2}
\nonumber\\
&&\hspace{2.0cm}-\frac{1}{2k_1}\log\frac{(4\mu^2-k_1^2)(k_1-k_2)^2}{4k_1^4(k_1+k_2)^2}+\frac{\eta_2^{(0)}}{k_2}-\frac{\eta_1^{(0)}}{k_1}.
\eea
From the above, it is evident that the bond length of the fundamental SM is predominantly influenced by the arbitrary phase constants, $\eta_j^{(0)}$, $j=1,2$, and wave numbers, $k_j$, $j=1,2$. For a specific SM, the phase constants, $\eta_j^{(0)}$'s, are arbitrary whereas the wave numbers, $k_j$'s, get restricted as per the velocity resonance condition (\ref{30}). By utilizing the arbitrariness of the phase constants one can tune the bond length between the soliton atoms. We get a dissociated molecule state  by following the condition, $\eta_1^{(0)}\neq \eta_2^{(0)}$, with larger separation distance. Such a dissociated molecule state is demonstrated in Fig. \ref{fg4}(a), where the two sech-type soliton atoms propagate in parallel (with $\Delta x_{12}=0.67$ which is calculated from soliton 1 to soliton 2) without any distortion. To get this figure we fix the wave parameters as $k_1=-2.9998$, $k_2=-2.9999$, $\eta_1^{(0)}=0$, and $\eta_2^{(0)}=2$.  Further, while considering $\eta_1^{(0)}=\eta_2^{(0)}$ or $\eta_1^{(0)}=\eta_2^{(0)}=0$, along with $k_{1}=-3.99985$ and $k_2=-3.9999$, one can observe that the bond length $\Delta x_{21}$ approaches zero value. In this case one obtains a synthetic molecule state that consists of two table-top soliton atoms with a minimal bond length $\Delta x_{21}=0.05$. A space-time plot corresponding to this molecule state (Fig. \ref{fg4}(b)) is illustrated in Fig. \ref{fg6}(a), where the relative separation between the atoms $S_2$ and $S_1$ is indicated as $D_{21}$. From Figs. \ref{fg4}(b) and its corresponding space-time plot \ref{fg6}(a) one can notice that no oscillation persists in the overlapping region. This special feature of the table-top synthetic molecule state makes it as distinct with a soliton molecule of the nonlinear Schr\"odinger equation, where quasi-periodic or periodic oscillations always appear in the coalescence region \cite{khawaja1,gelash,zakharov}. Therefore, based on the Gardner equation approach, one can confirm that for suitable choice of parameters constructive and destructive interference patterns do not occur in the soliton molecule of the FPUT lattice (\ref{fput}) with both cubic and quadratic nonlinear nearest neighbor interactions. To the best of our knowledge, the existence of table-top SM, in the FPUT lattice (\ref{fput}), has not been reported earlier in the literature. We note that one has to maintain a minimal spatial separation, though the phase constants are arbitrary,  between the atoms and it should be comparable with the full width at half maximum of a soliton atom.   
\begin{figure}[]
	\centering
	\includegraphics[width=0.9\linewidth]{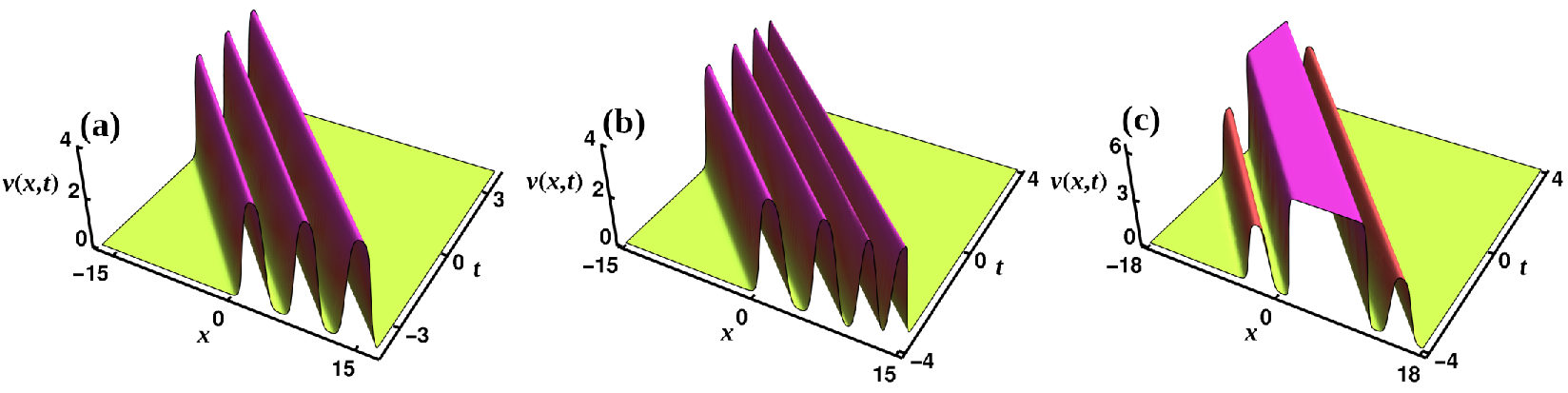}
	\caption{A synthetic tri-table-top soliton molecule is displayed in Fig. (a), in which the phase constants, $\eta_j^{(0)}$, $j=1,2,3$, obey $\eta_1^{(0)}=\eta_2^{(0)}=\eta_3^{(0)}$. A quad-soliton molecule is illustrated in Fig. (b) with the condition $\eta_1^{(0)}=\eta_2^{(0)}=\eta_3^{(0)}=\eta_4^{(0)}$. In Fig. (c), we display the formation of a mega table-top soliton in the coalescence region of quad-soliton molecule.    }
	\label{fg5}
\end{figure} 
\begin{figure}[]
	\centering
	\includegraphics[width=0.8\linewidth]{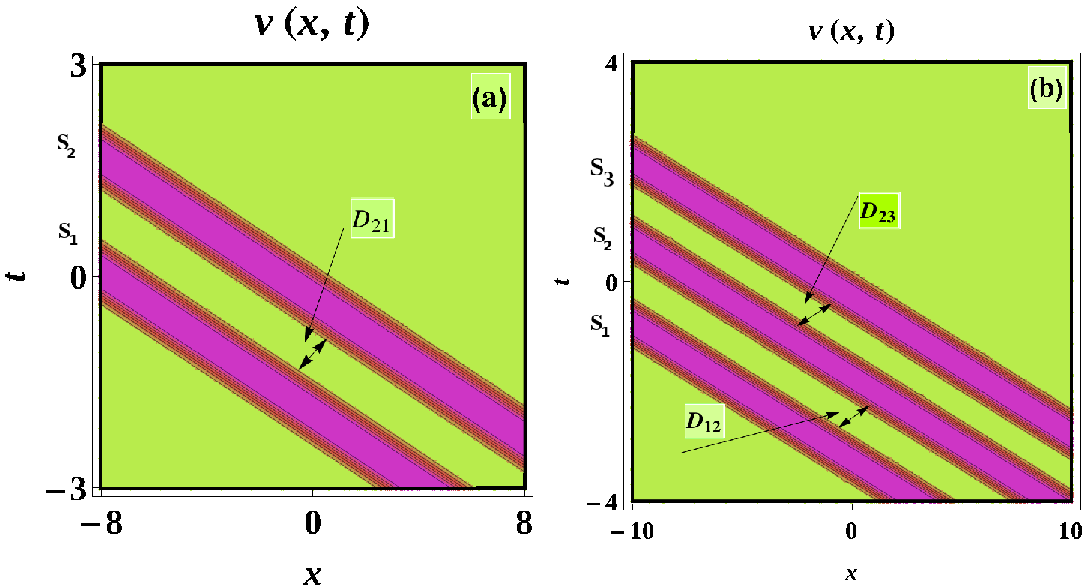}
	\caption{The space-time plot corresponding to Fig. \ref{fg4}(b) is illustrated in Fig. (a), where the bond length ($\Delta x_{21}$) between the atoms $S_2$ and $S_1$ is represented as $D_{21}$. We have drawn a contour plot corresponding to Fig. \ref{fg5}(a) in Fig. (b), in which the bond lengths ($\Delta x_{12}$ and $\Delta x_{23}$) between the soliton pairs $S_1$ and $S_2$, and $S_2$ and $S_3$ are represented as $D_{12}$ and $D_{23}$, respectively. }
	\label{fg6}
\end{figure} 

Besides the bi-soliton molecule, we also come across another special kind of soliton structure which is a breathing complexiton state when the wave number resonance condition \cite{wxma,alrazi1,alrazi2}, $k_2=k_1^*$ (that is $k_2=k_{2R}+ik_{2I}=k_1^*=k_{1R}-ik_{1I}$), is considered on the two-soliton solution (\ref{20a})-(\ref{20b}) of the Gardner equation (\ref{1}). 
Note that we have here relaxed the condition that $k_j$'s are always real so that the dynamical variable $v(x,t)$ is always real. An interesting aspect of the complexiton is that it exhibits breathing behavior, in both $x$ and $t$-directions, as it is depicted in Fig. \ref{fg4}(c). One can make an oscillation pattern along the spatial direction alone by setting a condition $k_{1I}^2-3k_{1R}^2=0$. In this case, the period of oscillation along the $x$ axis is $T=2\pi/k_{1I}$. We note that for the choice $k_{1I}=0$ the breathing complexiton state becomes a fundamental soliton. We remark that the complexiton displayed in Fig. \ref{fg4}(c) is drawn for $|v|$. However, the parameter values chosen by us still satisfy the Gardner equation (\ref{1}).              

Further, we obtain the tri-soliton molecular state from the three-soliton solution (\ref{21a})-(\ref{21b}) of Eq. (\ref{1}). The formation of such a three-soliton molecule can occur when the resonance condition  
\begin{equation}
	\frac{\omega_i}{\omega_j}=\frac{k_i^3}{k_j^3}~~ \text{or}~~\frac{v_i}{v_j}=\frac{k_i^2}{k_j^2}\rightarrow 1, ~i,j=1,2,3,\label{31}
\end{equation}
is satisfied by the wave numbers $k_j$, $j=1,2,3$. This condition induces a single molecular structure that contains three co-propagating solitons. This tri-molecular structure is displayed in Fig. \ref{fg5}(a), in which three table-top solitons are moving parallelly without any change in shapes. From this figure and its corresponding contour Fig. \ref{fg6}(b) one can confirm that the behviour of the tri-soliton molecule is qualitatively the same as that of the table-top bi-soliton molecule. To characterize the bond lengths between the soliton atoms in the present case we have calculated the spatial separation distances between the atoms as
\bes
\bea
\Delta x_{12}&=&\frac{1}{2k_1}\log\frac{(4\mu^2-k_1^2)(k_1-k_2)^2(k_1-k_3)^2}{4k_1^4(k_1+k_2)^2(k_1+k_3)^2}+\frac{\eta_1^{(0)}}{k_1}\nonumber\\
&&\hspace{-0.2cm}-\frac{\eta_2^{(0)}}{k_2}-\frac{1}{2k_2}\log\frac{(4\mu^2-k_2^2)(k_1-k_2)^2(k_2-k_3)^2}{4k_2^4(k_1+k_2)^2(k_2+k_3)^2},\\
\Delta x_{23}&=&\frac{1}{2k_2}\log\frac{(4\mu^2-k_2^2)(k_1-k_2)^2(k_2-k_3)^2}{4k_2^4(k_1+k_2)^2(k_2+k_3)^2}+\frac{\eta_2^{(0)}}{k_2}\nonumber\\
&&\hspace{-0.2cm}-\frac{\eta_3^{(0)}}{k_3}-\frac{1}{2k_3}\log\frac{(4\mu^2-k_3^2)(k_1-k_3)^2(k_2-k_3)^2}{4k_3^4(k_1+k_3)^2(k_2+k_3)^2}.
\eea\ees
Here, the relative spatial separation between the atoms $S_1$ and $S_2$ ($\Delta x_{12}$) and $S_2$ and $S_3$ ($\Delta x_{23}$) are indicated as $D_{12}$, and $D_{23}$, respectively, in Fig. \ref{fg6}(b). As it was mentioned before, to observe a dissociated tri-soliton molecule one has to set the condition, $\eta_1^{(0)}\neq \eta_2^{(0)}\neq \eta_3^{(0)}$, in which these  real constants are arbitrary. In this situation, the constituents are independent. However, as demonstrated in Fig. \ref{fg5}(a), a synthesis of table-top tri-soliton molecule occurs when we consider $\eta_1^{(0)}=\eta_2^{(0)}=\eta_3^{(0)}=0$ along with the wave numbers, $k_1=-3.9997$, $k_2=-3.9999$, $k_3=-3.9998$, obeying the condition (\ref{31}). 

A quad-molecular structure with homogeneous species of table-top solitons is also brought out from the four-soliton solution of the Gardner Eq. (\ref{1}). Such a co-moving soliton molecular structure is displayed in Fig. \ref{fg5}(b) by considering the velocity resonance condition, $\frac{\omega_i}{\omega_j}=\frac{k_i^3}{k_j^3}$, or $\frac{v_i}{v_j}=\frac{k_i^2}{k_j^2}\rightarrow 1$,  $i,j=1,2,3,4$ along with the restriction, $\eta_1^{(0)}=\eta_2^{(0)}=\eta_3^{(0)}=\eta_4^{(0)}=0$,  on the arbitrary constants $\eta_j^{(0)}$'s. To get the quad-SM (Fig. \ref{fg5}(b)), we fix the wave numbers as $k_1=-3.96$, $k_2=-3.998$, $k_3=-3.9997$, $k_3=-3.9999$. For brevity, we have omitted the details of bond lengths between constituents of quad-SM. However, in general, their analytical  forms can be identified from the four-soliton solution. Another interesting aspect of the quad-SM is that it leads to the formation of a mega table-top soliton in the coalescence region. This phenomenon occurs by the merging of two middle table-top solitons. We have demonstrated this interesting property of quad-soliton molecule in Fig. \ref{fg5}(c) for $k_1=-3.9997$, $k_2=-3.9999$, $k_3=-3.9998$, $k_3=-3.9996$, $\eta_1^{(0)}=\eta_4^{(0)}=0$, $\eta_2^{(0)}=2$, and $\eta_3^{(0)}=3$. We note that to obtain a more complex $N$-soliton molecular structure one has to consider the velocity resonance condition, $\omega_i/\omega_j=k_i^3/k_j^3$, or $v_i/v_j=k_i^2/k_j^2 \rightarrow 1$,  $i,j=1,2,3,...,N$. In this general case, one can have a more complicated form of molecule structure that consists of either pure dissociated type or pure synthetic type molecules or mixed type molecule states along with the formation of mega table-top solitons.           
\section{Collision and stability  properties of soliton molecules}
 In real physical systems many factors are responsible for the causes of perturbations which include gain, loss, impurities,  excitation of other kinds of waves in the same medium and other several physical factors. These perturbations, with enough strength, have the capability of destroying the SMs. As a consequence of this, the bonding between the constituents of the soliton compound will break and they tend to travel independently with their own identities. Therefore it is very important to study the stability of SMs. To examine this, we consider three situations. First, we allow the SM to interact with a fundamental  soliton of both sech and flat-top types as a weak perturbation. Then, next we investigate the collision of SM with two solitons, as strong perturbation. Finally, we consider a situation corresponding to the interaction of two fundamental bi-SMs. In the following we investigate each of the cases systematically with appropriate asymptotic analysis. 
 \begin{figure}[]
 	\centering
 	\includegraphics[width=0.78\linewidth]{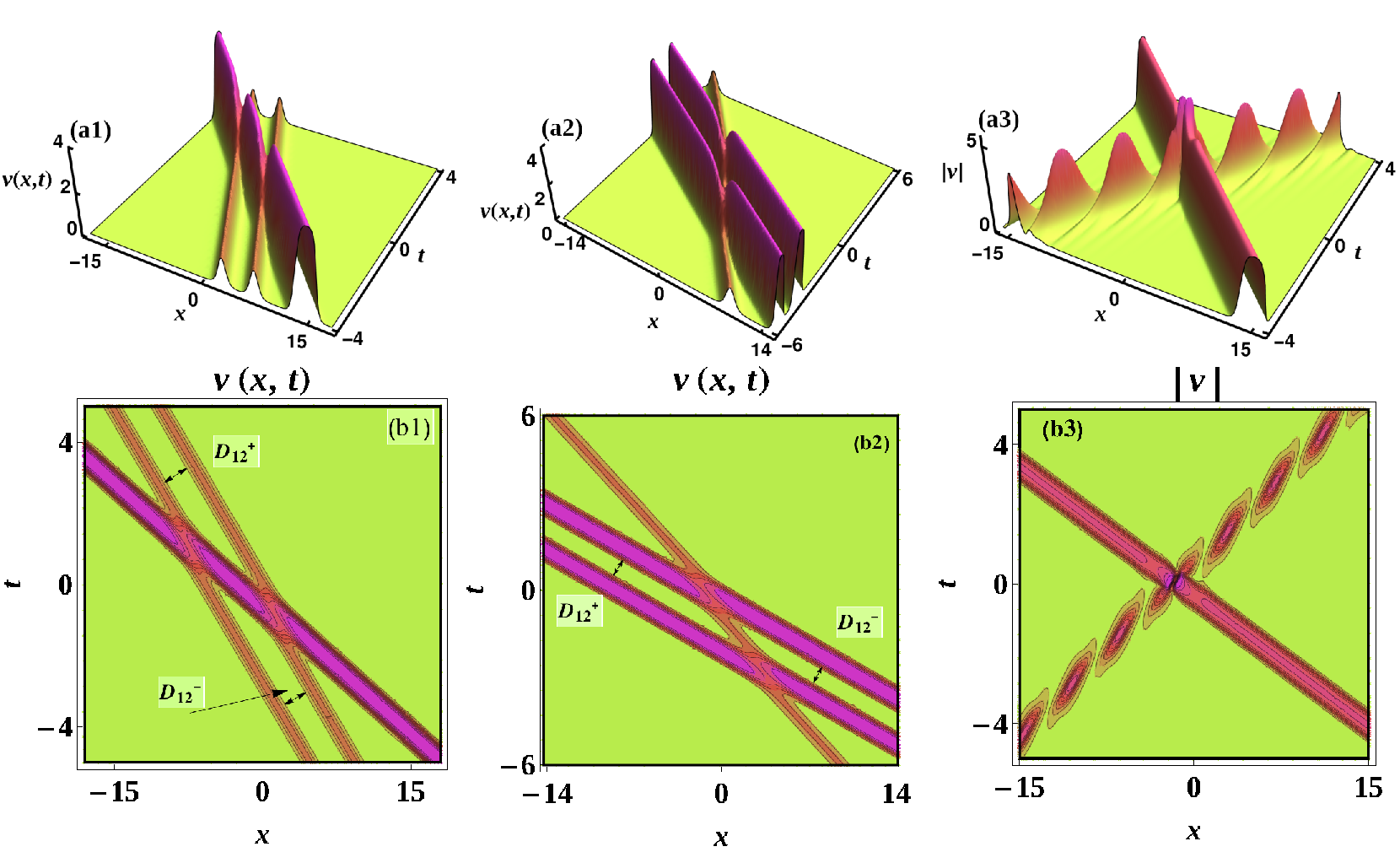}
 	\caption{The figures (a1)and (b1) illustrate an elastic interaction between a dissociated soliton molecule of sech-type and a flat-top soliton. The elastic collision between a synthetic table-top bi-soliton molecule and a sech-type soliton is depicted in Figs. (a2)-(b2). A collision plot corresponding to complexiton and a flat-top soliton is drawn in Figs. (a3) and (b3) for $k_1=0.65+2i$, $k_2=0.65-2i$, $k_3=-3.9999$, $\eta_1^{(0)}=0$, $\eta_2^{(0)}=1+i$, and $\eta_3^{(0)}=1$. }
 	\label{fg7}
 \end{figure}
\subsection{Collision between the bi-soliton molecule and a fundamental soliton: Asymptotic analysis}
To analyze the collision dynamics between a fundamental soliton molecule and a soliton we perform appropriate asymptotic analysis at the limits $t\rightarrow\pm \infty$. For this purpose, we assume the parametric choice as $k_{1,2}^2<k_3^2$, and $k_1^2\approx k_2^2$ ($v_1\approx v_2$).  In this situation, one can study the interaction of the dissociated (or synthetic) molecule of sech-type solitons with a table-top soliton. To find the asymptotic forms of the SM and a fundamental soliton we incorporate the asymptotic nature of the wave variables,   
$\eta_1=k_1(x+v_1t+\frac{\eta_1^{(0)}}{k_1})$, $\eta_2=k_2(x+v_2t+\frac{\eta_2^{(0)}}{k_2})$,
$\eta_3=k_3(x+v_3t+\frac{\eta_3^{(0)}}{k_3})$, $v_1=\frac{k_1^2}{4}$, $v_2=\frac{k_2^2}{4}$, $v_3=\frac{k_3^2}{4}$, in the three-soliton solution (\ref{21a})-(\ref{21b}). The asymptotic nature of the wave variables are obtained as (i) Molecule:  $\eta_{1,2}\approx 0$, $\eta_3\rightarrow \pm\infty$ as $t\rightarrow \mp \infty$, (ii) Soliton: $\eta_3\approx 0$, $\eta_{1,2}\rightarrow \pm \infty$, as $t\rightarrow \pm \infty$. By substituting these asymptotic behaviors of $\eta_j$'s in the solution (\ref{21a})-(\ref{21b}), we  deduce the following asymptotic forms for the soliton molecule as well as the one-soliton. \\\\
{\bf (a) Before collision: $t\rightarrow -\infty$}\\
{\bf Molecule: }$\eta_{1,2}\approx 0$, $\eta_3\rightarrow +\infty$\\
The exact form of bi-SM is deduced from the solution (\ref{21a})-(\ref{21b}) as 
\bea
&&\hspace{-0.4cm}v(x,t)=\frac{1}{D}\bigg(k_2^2(4\mu^2-k_1^2)^\frac{1}{2}\cosh(\eta_1+\vphi_1^-)+k_1^2(4\mu^2-k_2^2)^\frac{1}{2}\cosh(\eta_2+\vphi_2^-)\nonumber\\&&\hspace{1.0cm}-2\mu(k_1^2-k_2^2)\bigg),\label{33}\\
&&\hspace{-0.4cm}D=\frac{c_1}{2}(4\mu^2-k_1^2)^\frac{1}{2}(4\mu^2-k_2^2)^\frac{1}{2}\cosh(\eta_1+\eta_2+\vphi_1^-+\vphi_2^-)\nonumber\\&&\hspace{0.1cm}+\frac{1}{2c_1}(4\mu^2-k_1^2)^\frac{1}{2}(4\mu^2-k_2^2)^\frac{1}{2}\cosh(\eta_1-\eta_2+\vphi_1^--\vphi_2^-)\nonumber\\
&&\hspace{0.1cm}-2\mu(4\mu^2-k_1^2)^\frac{1}{2}\cosh(\eta_1+\vphi_1^-)-2\mu(4\mu^2-k_2^2)^\frac{1}{2}\cosh(\eta_2+\vphi_2^-)+c_2,\nonumber
\eea
where the phase terms corresponding to soliton atoms are  obtained as $\vphi_1^-=\frac{1}{2}\log\frac{(k_1-k_2)^2(k_1-k_3)^4(4\mu^2-k_1^2)}{4k_1^4(k_1+k_2)^2(k_1+k_3)^4}$, $\vphi_2^-=\frac{1}{2}\log\frac{(k_1-k_2)^2(k_2-k_3)^4(4\mu^2-k_2^2)}{4k_2^4(k_1+k_2)^2(k_2+k_3)^4}$. 
\\
{\bf Soliton:} $\eta_3\approx 0$, $\eta_{1,2}\rightarrow -\infty$\\
The asymptotic form of fundamental soliton is derived as follows: 
\bea
v(x,t)=\frac{k_3^2}{-2\mu+\sqrt{4\mu^2-k_3^2}\cosh(\eta_3+\phi_s^-)},\label{34}
\eea 
where $\phi_s^-=\frac{1}{2}\log\frac{4\mu^2-k_3^2}{4k_3^4}$.\\
{\bf (b) After collision: $t\rightarrow +\infty$}\\
{\bf Molecule:}  $\eta_{1,2}\approx 0$, $\eta_3\rightarrow -\infty$\\
The form of the fundamental SM is derived from the solution (\ref{21a})-(\ref{21b}) as 
\bea
&&\hspace{-0.4cm}v(x,t)=\frac{1}{D}\bigg(k_2^2(4\mu^2-k_1^2)^\frac{1}{2}\cosh(\eta_1+\vphi_1^+)+k_1^2(4\mu^2-k_2^2)^\frac{1}{2}\cosh(\eta_2+\vphi_2^+)\nonumber\\&&\hspace{1.0cm}-2\mu(k_1^2-k_2^2)\bigg),\label{35}\\
&&\hspace{-0.4cm}D=\frac{c_1}{2}(4\mu^2-k_1^2)^\frac{1}{2}(4\mu^2-k_2^2)^\frac{1}{2}\cosh(\eta_1+\eta_2+\vphi_1^++\vphi_2^+)\nonumber\\&&\hspace{0.1cm}+\frac{1}{2c_1}(4\mu^2-k_1^2)^\frac{1}{2}(4\mu^2-k_2^2)^\frac{1}{2}\cosh(\eta_1-\eta_2+\vphi_1^+-\vphi_2^+)\nonumber\\
&&\hspace{0.1cm}-2\mu(4\mu^2-k_1^2)^\frac{1}{2}\cosh(\eta_1+\vphi_1^+)-2\mu(4\mu^2-k_2^2)^\frac{1}{2}\cosh(\eta_2+\vphi_2^+)+c_2,\nonumber
\eea
where the phase terms of the soliton atoms are given by $\vphi_1^+=\frac{1}{2}\log\frac{(k_1-k_2)^2(4\mu^2-k_1^2)}{4k_1^4(k_1+k_2)^2}$, $\vphi_2^+=\frac{1}{2}\log\frac{(k_1-k_2)^2(4\mu^2-k_2^2)}{4k_2^4(k_1+k_2)^2}$.
\\
{\bf Soliton: }$\eta_3\approx 0$, $\eta_{1,2}\rightarrow 
+\infty$\\
The exact asymptotic form of one-soliton is deduced as
\bea
v(x,t)=\frac{k_3^2}{-2\mu+\sqrt{4\mu^2-k_3^2}\cosh(\eta_3+\phi_s^+)},\label{36}
\eea 
here, $\phi_s^+=\frac{1}{2}\log\frac{4\mu^2-k_3^2}{4k_3^4}+\frac{1}{2}\log\frac{(k_1-k_3)^4(k_2-k_3)^4}{(k_1+k_3)^4(k_2+k_3)^4}$. 

From the above, we identify that the phases of soliton atoms, after and before collisions, are related as follows: 
\bea
\vphi_1^+=\vphi_1^--\frac{1}{2}\log\frac{(k_1-k_3)^4}{(k_1+k_3)^4},~~~\vphi_2^+=\vphi_2^--\frac{1}{2}\log\frac{(k_2-k_3)^4}{(k_2+k_3)^4}.\label{37}
\eea
Similarly, we also find that the phase of fundamental soliton after and before collision is also related in the following way:  
\begin{equation}
\phi_s^+=\phi_s^-+\frac{1}{2}\log\frac{(k_1-k_3)^4(k_2-k_3)^4}{(k_1+k_3)^4(k_2+k_3)^4}. \label{38}
\end{equation}

The above asymptotic analysis clearly demonstrates that the structures (amplitude and width) of both the SM and the fundamental soliton are preserved throughout the collision process, except for the variations in the phase shifts. It means that the corresponding asymptotic expressions, in both the limits $t\rightarrow \pm \infty$, of them are invariant under collision, except for the changes in the phase terms. One can easily find that the  expressions (\ref{33})-(\ref{36}) substantiate this as true. Consequently, the collision scenario observed between the bi-SM and a fundamental soliton is an elastic one. However, one could ask the question: do the variations in phase shifts affect the dynamics of the molecule and soliton? As far as the fundamental soliton is concerned the change in phases only shift the position of the soliton after collision. On the other hand, the variations in the phase shifts of the soliton atoms directly influence the bond length of the bi-SM. Therefore, this implies that the change in the group phase shift (that is, the relative separation distance between the soliton atoms in a soliton molecule), in general, slightly distorts the bond-length of the bi-SM. In contrast, though we observe a difference in the group phase shift, the structure of the fundamental SM remains unchanged and it shows stable propagation throughout the evolution process as it is depicted in Figs. \ref{fg7}(a1) and \ref{fg7}(b1). The group phase shift of a bi-soliton molecule is specified by the relative separation distance between the constituent atoms. To find this we have calculated the spatial separation between the atoms before and after collision and they turn out to be             
\bes\bea
\Delta x_{12}^-~ (D_{12}^-)&=&\frac{\eta_2^{(0)}}{k_2}+\frac{1}{2k_2}\log\frac{(4\mu^2-k_2^2)(k_1-k_2)^2(k_2-k_3)^4}{4k_2^4(k_1+k_2)^2(k_2+k_3)^4}\nonumber\\
&&\hspace{-0.2cm}-\frac{1}{2k_1}\log\frac{(4\mu^2-k_1^2)(k_1-k_2)^2(k_1-k_3)^4}{4k_1^4(k_1+k_2)^2(k_1+k_3)^4}-\frac{\eta_1^{(0)}}{k_1}, \label{39a} \\
\Delta x_{12}^+~(D_{12}^+)&=&\frac{\eta_2^{(0)}}{k_2}+\frac{1}{2k_2}\log\frac{(4\mu^2-k_2^2)(k_1-k_2)^2}{4k_2^4(k_1+k_2)^2}\nonumber\\
&&-\frac{\eta_1^{(0)}}{k_1}-\frac{1}{2k_1}\log\frac{(4\mu^2-k_1^2)(k_1-k_2)^2}{4k_1^4(k_1+k_2)^2}. \label{39b}
\eea\ees
Here,  $\Delta x_{12}^-$ and $\Delta x_{12}^+$ denote the bond length between the atoms before and after collision, respectively. This implies that during the collision process the bond length of the bi-soliton molecule varies while it interacts with a fundamental soliton. We also calculate the total change in the bond length of the molecule during the entire collision process. It is obtained as
\bea
\Del \Phi=\Delta x_{12}^+-\Delta x_{12}^-=\frac{1}{2k_2}\log\frac{(k_2+k_3)^4}{(k_2-k_3)^4}-\frac{1}{2k_1}\log\frac{(k_1+k_3)^4}{(k_1-k_3)^4}.
\eea
From the above relation, we find that the presence of the wave number $k_3$, corresponding to the fundamental soliton, significantly affects the bond length between the soliton atoms. A typical elastic collision between a dissociated bi-SM of sech-type and a flat-top soliton is illustrated as an example in Fig. \ref{fg7}(a1) for the parameter values $k_1=-2.9$, $k_2=-2.92$, $k_3=-3.9999$, $\eta_1^{(0)}=2$, $\eta_2^{(0)}=0$, and $\eta_3^{(0)}=0$.
From this figure, one can confirm that the structure of the bi-SM is preserved under collision without any visible changes in bond length, apart from no structural change in the fundamental soliton profile. We verify the bond length variation using the formulae (\ref{39a})-(\ref{39b}) along with the parameter values corresponding to  Fig. \ref{fg7}(a1). By doing so, we identified the bond length of the SM corresponding to before and after collision as $\Delta x_{12}^-=0.68$, $\Delta x_{12}^+=0.67$, respectively. These values ensure that the bi-SM in Fig. \ref{fg7}(a1) experiences only a slight variation in the bond length after collision. Therefore, the entire molecular structure of sech-type continues to propagate in a stable manner even after collision with a fundamental table-top soliton. In addition to this, we also confirm this property by considering the interaction between a synthetic table-top bi-SM and a sech-type fundamental soliton. Such an elastic collision scenario is displayed in Figs. \ref{fg7}(a2)-(b2) for  $k_1=-2.9$, $k_2=-3.9999$, $k_3=-3.9998$, $\eta_1^{(0)}=0$, $\eta_2^{(0)}=0$, and $\eta_3^{(0)}=0$. These collision scenarios ensure that the soliton molecule in the FPUT lattice (\ref{1}) withstands the effect of weak perturbation and retains its structure. Besides the above, one can observe another interesting feature of the fundamental Gardner soliton from Figs. \ref{fg7}(a1) and (b1) which is that it undergoes two consecutive collisions with the two soliton atoms of the SM. As a result, the fundamental soliton  experiences two consecutive phase shifts. However, such successive collisions do not affect the soliton profile structure as this is evident from Figs. \ref{fg7}(a1) and (b1) and also from the asymptotic forms (\ref{34}) and (\ref{36}). We remark here that the existence of such a SM with bond length variation under collision has not been explained with analytical support in the earlier literature on SMs \cite{lou1}. For completeness, in Figs. \ref{fg7}(a3)-(b3), we also illustrate the collision between a complexiton and a flat-top soliton. From these figures, one can confirm that their structures remain unchanged during the entire collision scenario.        
\begin{figure}[]
	\centering
	\includegraphics[width=0.86\linewidth]{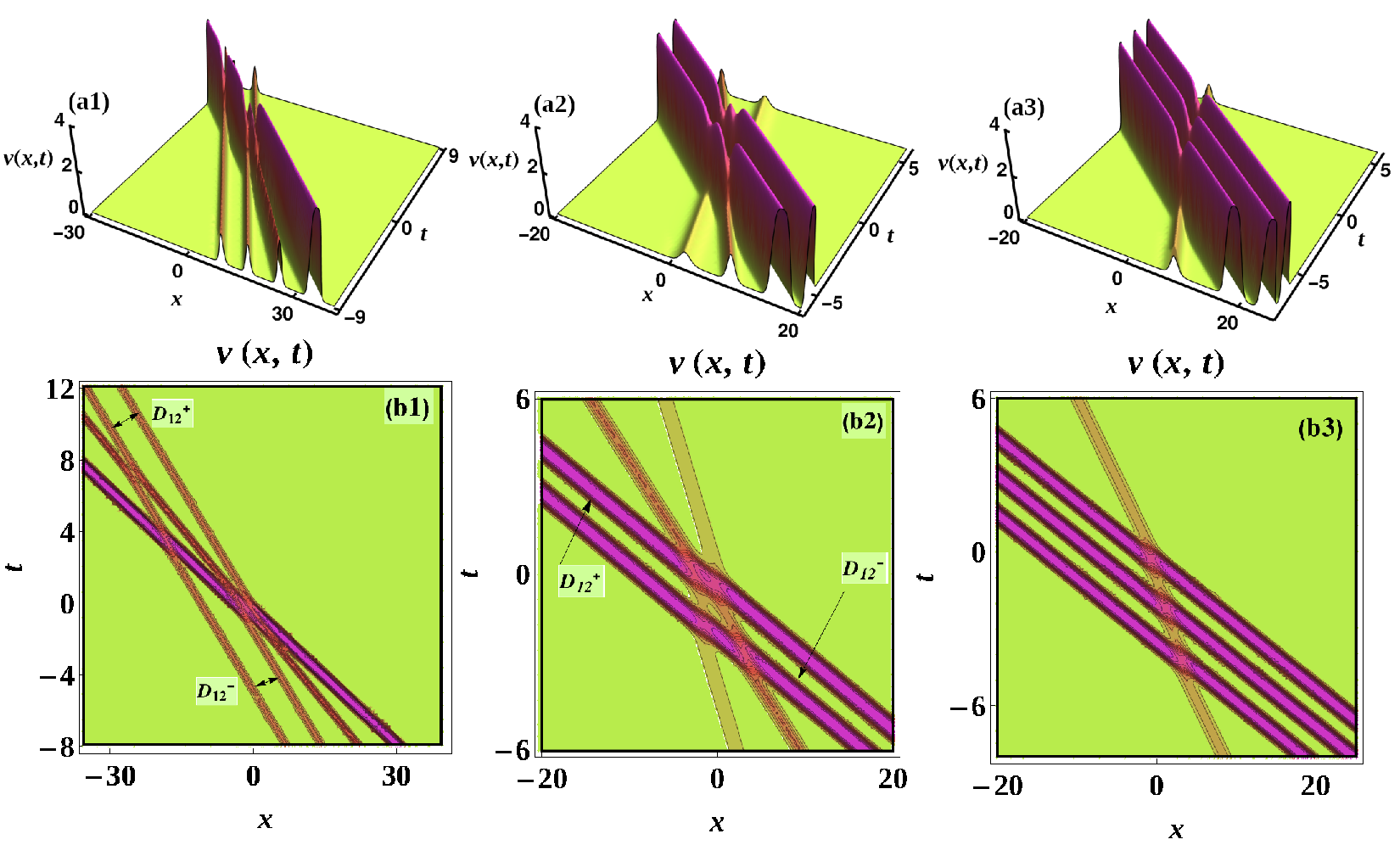}
	\caption{The figures (a1) and (b1) demonstrate an elastic interaction between a dissociated soliton molecule of sech-type and two fundamental solitons. In Figs. (a2) and (b2) we illustrate an elastic interaction between a synthetic soliton molecule of table-top type and two fundamental solitons. The Figs. (a3) and (b3) denote an elastic collision between a tri-soliton molecule and a fundamental soliton. }
	\label{fg8}
\end{figure}
\subsection{Collision between the soliton molecule and two solitons: Asymptotic analysis}
Next, we consider a situation where the bi-soliton molecule experiences a strong perturbation when it collides with two fundamental solitons. We investigate this situation by performing an asymptotic analysis along with the parametric choice $k_{1,2}^2<k_3^2<k_4^2$. Under this choice, one has to consider the asymptotic nature of the wave variables,   
$\eta_j=k_j(x+v_jt+\frac{\eta_j^{(0)}}{k_j})$,  $v_j=\frac{k_j^2}{4}$, $j=1,2,3,4$, in the four-soliton solution of the Gardner Eq. (\ref{1}). By doing so, the asymptotic nature of the wave variables are obtained as (i) Molecule:  $\eta_{1,2}\approx 0$, $\eta_3\rightarrow \pm\infty$, $\eta_4\rightarrow \pm\infty$ as $t\rightarrow \mp \infty$, (ii) Soliton 1: $\eta_3\approx 0$, $\eta_{1,2}\rightarrow \pm \infty$, $\eta_4\rightarrow \mp\infty$ as $t\rightarrow \pm \infty$, and (iii) Soliton 2: $\eta_4\approx 0$, $\eta_{1,2}\rightarrow \pm \infty$, $\eta_3\rightarrow \pm\infty$ as $t\rightarrow \pm \infty$. By considering these asymptotic behaviors of $\eta_j$'s in the four-soliton solution of the Gardner equation we  deduce the following asymptotic forms for the bi-SM and two individual solitons. \\
{\bf (a) Before collision:} $t\rightarrow -\infty$\\
{\bf Molecule: }$\eta_{1,2}\approx 0$, $\eta_{3,4}\rightarrow +\infty$\\
The asymptotic expression of the bi-soliton molecule is obtained as 
\bea
&&\hspace{-0.4cm}v(x,t)=\frac{1}{D}\bigg(k_2^2(4\mu^2-k_1^2)^\frac{1}{2}\cosh(\eta_1+\vphi_1^-)+k_1^2(4\mu^2-k_2^2)^\frac{1}{2}\cosh(\eta_2+\vphi_2^-)\nonumber\\&&\hspace{1.0cm}-2\mu(k_1^2-k_2^2)\bigg),\eea\bea
&&\hspace{-0.4cm}D=\frac{c_1}{2}(4\mu^2-k_1^2)^\frac{1}{2}(4\mu^2-k_2^2)^\frac{1}{2}\cosh(\eta_1+\eta_2+\vphi_1^-+\vphi_2^-)\nonumber\\&&\hspace{0.1cm}+\frac{1}{2c_1}(4\mu^2-k_1^2)^\frac{1}{2}(4\mu^2-k_2^2)^\frac{1}{2}\cosh(\eta_1-\eta_2+\vphi_1^--\vphi_2^-)\nonumber\\
&&\hspace{0.1cm}-2\mu(4\mu^2-k_1^2)^\frac{1}{2}\cosh(\eta_1+\vphi_1^-)-2\mu(4\mu^2-k_2^2)^\frac{1}{2}\cosh(\eta_2+\vphi_2^-)+c_2,\nonumber
\eea
where the phase terms are identified as  $\vphi_1^-=\frac{1}{2}\log\frac{(k_1-k_2)^2(k_1-k_3)^4(k_1-k_4)^4(4\mu^2-k_1^2)}{4k_1^4(k_1+k_2)^2(k_1+k_3)^4(k_1+k_4)^4}$, and $\vphi_2^-=\frac{1}{2}\log\frac{(k_1-k_2)^2(k_2-k_3)^4(k_2-k_4)^4(4\mu^2-k_2^2)}{4k_2^4(k_1+k_2)^2(k_2+k_3)^4(k_2+k_4)^4}$.\\
{\bf Soliton 1:} $\eta_3\approx 0$, $\eta_{1,2}\rightarrow -\infty$, $\eta_4\rightarrow +\infty$ \\
The asymptotic form of the soliton 1 is deduced from the four-soliton solution as 
\bea
v(x,t)=\frac{k_3^2}{-2\mu+\sqrt{4\mu^2-k_3^2}\cosh(\eta_3+\vphi_3^-)},
\eea 
where $\vphi_3^-=\frac{1}{2}\log\frac{(k_3-k_4)^4(4\mu^2-k_3^2)}{4k_3^4(k_3+k_4)^4}$.\\
{\bf Soliton 2:} $\eta_4\approx 0$, $\eta_{1,2}\rightarrow - \infty$, $\eta_3\rightarrow -\infty$\\
The form of soliton 2 is derived as
\bea
v(x,t)=\frac{k_4^2}{-2\mu+\sqrt{4\mu^2-k_4^2}\cosh(\eta_4+\vphi_4^-)}.
\eea 
Here the phase of the soliton $2$ is obtained as $\vphi_4^-=\frac{1}{2}\log\frac{(4\mu^2-k_4^2)}{4k_4^4}$.\\
{\bf (b) After collision:} $t\rightarrow +\infty$\\
{\bf Molecule}:$\eta_{1,2}\approx 0$, $\eta_{3,4}\rightarrow -\infty$\\
The asymptotic form of the bi-SM is found to be
\bea
&&\hspace{-0.4cm}v(x,t)=\frac{1}{D}\bigg(k_2^2(4\mu^2-k_1^2)^\frac{1}{2}\cosh(\eta_1+\vphi_1^+)+k_1^2(4\mu^2-k_2^2)^\frac{1}{2}\cosh(\eta_2+\vphi_2^+)\nonumber\\&&\hspace{1.0cm}-2\mu(k_1^2-k_2^2)\bigg),\eea\bea
&&\hspace{-0.4cm}D=\frac{c_1}{2}(4\mu^2-k_1^2)^\frac{1}{2}(4\mu^2-k_2^2)^\frac{1}{2}\cosh(\eta_1+\eta_2+\vphi_1^++\vphi_2^+)\nonumber\\&&\hspace{0.1cm}+\frac{1}{2c_1}(4\mu^2-k_1^2)^\frac{1}{2}(4\mu^2-k_2^2)^\frac{1}{2}\cosh(\eta_1-\eta_2+\vphi_1^+-\vphi_2^+)\nonumber\\
&&\hspace{0.1cm}-2\mu(4\mu^2-k_1^2)^\frac{1}{2}\cosh(\eta_1+\vphi_1^+)-2\mu(4\mu^2-k_2^2)^\frac{1}{2}\cosh(\eta_2+\vphi_2^+)+c_2,\nonumber
\eea
where the phase terms $\vphi_j^+$, $j=1,2$, are identified as $\vphi_1^+=\frac{1}{2}\log\frac{(k_1-k_2)^2(4\mu^2-k_1^2)}{4k_1^4(k_1+k_2)^2}$, $\vphi_2^+=\frac{1}{2}\log\frac{(k_1-k_2)^2(4\mu^2-k_2^2)}{4k_2^4(k_1+k_2)^2}$.\\
{\bf Soliton 1:}$\eta_3\approx 0$, $\eta_{1,2}\rightarrow +\infty$, $\eta_4\rightarrow -\infty$ \\
In this limit, the asymptotic expression of the soliton 1 is found to be
\bea
v(x,t)=\frac{k_3^2}{-2\mu+\sqrt{4\mu^2-k_3^2}\cosh(\eta_3+\vphi_3^+)}
\eea 
where $\vphi_3^+=\frac{1}{2}\log\frac{(k_1-k_3)^4(k_2-k_3)^4(4\mu^2-k_3^2)}{4k_3^4(k_1+k_3)^4(k_2+k_3)^4}=\vphi_3^-+\frac{1}{2}\log\frac{(k_1-k_3)^4(k_2-k_3)^4(k_3+k_4)^4}{(k_1+k_3)^4(k_2+k_3)^4(k_3-k_4)^4}$.\\
{\bf Soliton 2: }$\eta_4\approx 0$, $\eta_{1,2}\rightarrow +\infty$, $\eta_3\rightarrow +\infty$\\
The form of soliton 2 is deduced, by considering the dominant terms from the four-soliton solution, as follows:
\bea
v(x,t)=\frac{k_4^2}{-2\mu+\sqrt{4\mu^2-k_4^2}\cosh(\eta_4+\vphi_4^+)},
\eea 
where the phase term is given by  $\vphi_4^+=\frac{1}{2}\log\frac{(4\mu^2-k_4^2)}{4k_4^4}+\frac{1}{2}\log\frac{(k_1-k_4)^4(k_2-k_4)^4(k_3-k_4)^4}{(k_1+k_4)^4(k_2+k_4)^4(k_3+k_4)^4}$. The latter expression can be rewritten as  $\vphi_4^+=\vphi_4^-+\frac{1}{2}\log\frac{(k_1-k_4)^4(k_2-k_4)^4(k_3-k_4)^4}{(k_1+k_4)^4(k_2+k_4)^4(k_3+k_4)^4}$. From the above, we find that the phases of soliton atoms before and after collision are related as follows: 
\bes\begin{eqnarray}
\vphi_1^-=\vphi_1^++\frac{1}{2}\log\frac{(k_1-k_3)^4(k_1-k_4)^2}{(k_1+k_3)^4(k_1+k_4)^4},\\
\vphi_2^-=\vphi_2^++\frac{1}{2}\log\frac{(k_2-k_3)^4(k_2-k_4)^4}{(k_2+k_3)^4(k_2+k_4)^4}.
\end{eqnarray}\ees

The above results of asymptotic analysis again confirm that the nature of collision between the bi-soliton molecule and two individual Gardner solitons is an elastic  one since their structures remain the same throughout the evolution process. However, as it was mentioned in the previous case,  in the present situation also the two individual solitons undergo two consecutive collisions with the soliton atoms. As a consequence of this, the two solitons experience two consecutive phase shifts whereas the bi-soliton molecule undergoes a group phase shift.  Due to this fact, during the collision scenario the positions of the two fundamental solitons are shifted to new positions from their respective original positions. On the other hand, the occurrence of group phase shift, due to two consecutive collisions with the fundamental soliton, in the SM leads to a drastic variation in the bond length. This is reflected as a deformation the structure of the SM. The variation in the bond length is calculated as 
\bes\bea
&&\hspace{-1.0cm}\Delta x_{12}^-~ (D_{12}^-)=\frac{\eta_2^{(0)}}{k_2}+\frac{1}{2k_2}\log\frac{(4\mu^2-k_2^2)(k_1-k_2)^2(k_2-k_3)^4(k_2-k_4)^4}{4k_2^4(k_1+k_2)^2(k_2+k_3)^4(k_2+k_4)^4}\nonumber\\
&&~~~~~-\frac{1}{2k_1}\log\frac{(4\mu^2-k_1^2)(k_1-k_2)^2(k_1-k_3)^4(k_1-k_4)^4}{4k_1^4(k_1+k_2)^2(k_1+k_3)^4(k_1+k_4)^4}-\frac{\eta_1^{(0)}}{k_1},~~ \label{48a} \\
&&\hspace{-1.0cm}\Delta x_{12}^+~(D_{12}^+)=\frac{\eta_2^{(0)}}{k_2}+\frac{1}{2k_2}\log\frac{(4\mu^2-k_2^2)(k_1-k_2)^2}{4k_2^4(k_1+k_2)^2}\nonumber\\
&&~~~~~~~~-\frac{\eta_1^{(0)}}{k_1}-\frac{1}{2k_1}\log\frac{(4\mu^2-k_1^2)(k_1-k_2)^2}{4k_1^4(k_1+k_2)^2}. \label{48b}
\eea\ees
 The above relations show the possibility of occurrence of bond length deformation in a molecule while it interacts with two fundamental solitons. We have also calculated the total change in the bond length of a bi-soliton molecule during the collision and it reads as
\bea
\Del \Phi&=&\Delta x_{12}^+-\Delta x_{12}^-\nonumber\\&=&\frac{1}{2k_2}\log\frac{(k_2+k_3)^4(k_2+k_4)^4}{(k_2-k_3)^4(k_2-k_4)^4}-\frac{1}{2k_1}\log\frac{(k_1+k_3)^4(k_1+k_4)^4}{(k_1-k_3)^4(k_1-k_4)^4}.
\eea     
From the above calculation on bond length, we observe that the bond length variation may influence the structure of the fundamental SM. However, as we describe below, the entire molecular structure propagates in a stable way without any effects on the amplitudes and widths of its constituent atoms.  

As an example, a typical collision between a bi-SM of dissociate type and two fundamental solitons is displayed in Figs. \ref{fg8}(a1) and \ref{fg8}(b1), where  a sech-type bi-SM  interacts with a sech- and a table-top type solitons. To bring out this collision picture, the parameter values are fixed as  $k_1=-2.9991$, $k_2=-2.9993$, $k_3=-3.5$, $k_4=-3.9999$, $\eta_1^{(0)}=1$, $\eta_2^{(0)}=0$, $\eta_3^{(0)}=1$, and $\eta_4^{(0)}=2$. From these figures, we observe that the two solitons undergo two successive phase shifts, which can be calculated in general from the asymptotic forms. However, their structures remain unchanged during the collision process. Then, as far as the SM is concerned,  the variation in bond length, before and after collision, is calculated from the formulae (\ref{48a})-(\ref{48b}) as $\Delta x_{12}^-=-0.5005$, $\Delta x_{12}^+=0.3333$, respectively. These bond length values show that a slight deformation does occur in the structure of the bi-soliton molecule. Such a deformation leads to further dissociation of the soliton atoms from the already dissociated soliton compound. However, as it is evident from Figs.  \ref{fg8}(a1) and (b1), the dissociated bi-soliton molecule evolves in a stable manner even after experiencing the strong collisions with the two solitons. This further ensures the elastic nature of the interaction between a fundamental SM and the two solitons. To further substantiate this, we have also illustrated the interaction between a synthetic molecule of table-top soliton type and two solitons in Figs. \ref{fg8}(a2) and (b2). In this situation also we observe a slight variation in the bond-length of this synthetic molecule. However, both the molecule and the two solitons   
undergo  shape preserving collision thereby confirming an elastic collision. Further, to display the stability of a higher-order soliton molecule, we consider the collision between a tri-soliton molecule and a fundamental soliton in Figs. \ref{fg8}(a3)-(c3) with the parameter values  
$k_1=-2.5$, $k_2=-3.99975$, $k_3=-3.9998$, $k_4=-3.9999$, $\eta_j^{(0)}=0$, $j=1,2,3,4$. The figure clearly indicates that the tri-soliton molecule is also not prone to the perturbation and it is indeed stable against soliton collision. In this situation, the fundamental soliton experiences three successive phase shifts. From the above discussion, we re-ensure that the soliton molecule in the Gardner equation or FPUT lattice (\ref{1}) is stable against strong perturbation as well.       
\subsection{Collision among the two soliton molecules: Asymptotic analysis}
Now, we consider a case in which a dissociated bi-soliton molecule interacts with a synthetic bi-soliton molecule. For this purpose the parametric choice, $k_4^2>k_3^2>k_2^2>k_1^2$ (but $k_1^2\approx k_2^2$ , $k_3^2\approx k_4^2$), is considered for performing asymptotic analysis. Under this choice and at the asymptotic limits $t\rightarrow \pm \infty$, we find that the wave variables $\eta_j$'s  behave asymptotically as (i) Molecule 1:  $\eta_{1,2}\approx 0$, $\eta_{3,4}\rightarrow \pm\infty$, as $t\rightarrow \mp \infty$,   (ii) Molecule 2:  $\eta_{3,4}\approx 0$, $\eta_{1,2}\rightarrow \mp\infty$ as $t\rightarrow \mp \infty$. To deduce the asymptotic expressions corresponding to the two molecules, we incorporate the latter behavior of the wave variables in the four-soliton solution of the Gardner equation (\ref{1}). By doing so, we arrive the following asymptotic forms of SMs.       
\\
{\bf (a) Before collision: $t\rightarrow -\infty$}\\
{\bf Molecule  1: }$\eta_{1,2}\approx 0$, $\eta_{3,4}\rightarrow +\infty$\\
In this limit, we have deduced the following asymptotic form for the soliton molecule 1. It reads as 
\bea
&&\hspace{-0.4cm}v(x,t)=\frac{1}{D}\bigg(k_2^2(4\mu^2-k_1^2)^\frac{1}{2}\cosh(\eta_1+\vphi_1^-)+k_1^2(4\mu^2-k_2^2)^\frac{1}{2}\cosh(\eta_2+\vphi_2^-)\nonumber\\&&\hspace{1.0cm}-2\mu(k_1^2-k_2^2)\bigg),\label{50}\\
&&\hspace{-0.4cm}D=\frac{c_1}{2}(4\mu^2-k_1^2)^\frac{1}{2}(4\mu^2-k_2^2)^\frac{1}{2}\cosh(\eta_1+\eta_2+\vphi_1^-+\vphi_2^-)\nonumber\\&&\hspace{0.1cm}+\frac{1}{2c_1}(4\mu^2-k_1^2)^\frac{1}{2}(4\mu^2-k_2^2)^\frac{1}{2}\cosh(\eta_1-\eta_2+\vphi_1^--\vphi_2^-)\nonumber\\
&&\hspace{0.1cm}-2\mu(4\mu^2-k_1^2)^\frac{1}{2}\cosh(\eta_1+\vphi_1^-)-2\mu(4\mu^2-k_2^2)^\frac{1}{2}\cosh(\eta_2+\vphi_2^-)+c_2,\nonumber
\eea
where the phase terms are given by $\vphi_1^-=\frac{1}{2}\log\frac{(k_1-k_2)^2(k_1-k_3)^4(k_1-k_4)^4(4\mu^2-k_1^2)}{4k_1^4(k_1+k_2)^2(k_1+k_3)^4(k_1+k_4)^4}$, $\vphi_2^-=\frac{1}{2}\log\frac{(k_1-k_2)^2(k_2-k_3)^4(k_2-k_4)^4(4\mu^2-k_2^2)}{4k_2^4(k_1+k_2)^2(k_2+k_3)^4(k_2+k_4)^4}$.\\
{\bf Molecule  2: }$\eta_{3,4}\approx 0$, $\eta_{1,2}\rightarrow -\infty$ \\
The following asymptotic expression is deduced for the molecule 2 from the four-soliton solution of Eq. (\ref{1}):
\bea
&&\hspace{-0.4cm}v(x,t)=\frac{1}{D}\bigg(k_4^2(4\mu^2-k_3^2)^\frac{1}{2}\cosh(\eta_3+\vphi_3^-)+k_3^2(4\mu^2-k_4^2)^\frac{1}{2}\cosh(\eta_4+\vphi_4^-)\nonumber\\&&\hspace{1.0cm}-2\mu(k_3^2-k_4^2)\bigg),\\
&&\hspace{-0.4cm}D=\frac{c_3}{2}(4\mu^2-k_3^2)^\frac{1}{2}(4\mu^2-k_4^2)^\frac{1}{2}\cosh(\eta_3+\eta_4+\vphi_3^-+\vphi_4^-)\nonumber\\&&\hspace{0.1cm}+\frac{1}{2c_3}(4\mu^2-k_3^2)^\frac{1}{2}(4\mu^2-k_4^2)^\frac{1}{2}\cosh(\eta_3-\eta_4+\vphi_3^--\vphi_4^-)\nonumber\\
&&\hspace{0.1cm}-2\mu(4\mu^2-k_3^2)^\frac{1}{2}\cosh(\eta_3+\vphi_3^-)-2\mu(4\mu^2-k_4^2)^\frac{1}{2}\cosh(\eta_4+\vphi_4^-)+c_4.\nonumber
\eea
Here the phase terms are identified as $\vphi_3^-=\frac{1}{2}\log\frac{(k_3-k_4)^2(4\mu^2-k_3^2)}{4k_3^4(k_3+k_4)^2}$,\\ $\vphi_4^-=\frac{1}{2}\log\frac{(k_3-k_4)^2(4\mu^2-k_4^2)}{4k_4^4(k_3+k_4)^2}$, and the constants are obtained as $c_3=\frac{(k_3-k_4)}{(k_3+k_4)}$, $c_4=\frac{2[2\mu^2(k_3^2+k_4^2)-k_3^2k_4^2]}{k_3^2-k_4^2}$.\\
{\bf (b) After collision: }$t\rightarrow +\infty$\\
{\bf Molecule  1: } $\eta_{1,2}\approx 0$, $\eta_{3,4}\rightarrow -\infty$\\
In this limit, the form corresponding to the molecule 1 is obtained as follows: 
\bea
&&\hspace{-0.4cm}v(x,t)=\frac{1}{D}\bigg(k_2^2(4\mu^2-k_1^2)^\frac{1}{2}\cosh(\eta_1+\vphi_1^+)+k_1^2(4\mu^2-k_2^2)^\frac{1}{2}\cosh(\eta_2+\vphi_2^+)\nonumber\\&&\hspace{1.0cm}-2\mu(k_1^2-k_2^2)\bigg),\\
&&\hspace{-0.4cm}D=\frac{c_1}{2}(4\mu^2-k_1^2)^\frac{1}{2}(4\mu^2-k_2^2)^\frac{1}{2}\cosh(\eta_1+\eta_2+\vphi_1^++\vphi_2^+)\nonumber\\&&\hspace{0.1cm}+\frac{1}{2c_1}(4\mu^2-k_1^2)^\frac{1}{2}(4\mu^2-k_2^2)^\frac{1}{2}\cosh(\eta_1-\eta_2+\vphi_1^+-\vphi_2^+)\nonumber\\
&&\hspace{0.1cm}-2\mu(4\mu^2-k_1^2)^\frac{1}{2}\cosh(\eta_1+\vphi_1^+)-2\mu(4\mu^2-k_2^2)^\frac{1}{2}\cosh(\eta_2+\vphi_2^+)+c_2.\nonumber
\eea
In the above, the phase terms are obtained as  $\vphi_1^+=\frac{1}{2}\log\frac{(k_1-k_2)^2(4\mu^2-k_1^2)}{4k_1^4(k_1+k_2)^2}$, $\vphi_2^+=\frac{1}{2}\log\frac{(k_1-k_2)^2(4\mu^2-k_2^2)}{4k_2^4(k_1+k_2)^2}$.\\
{\bf Molecule  2:} $\eta_{3,4}\approx 0$, $\eta_{1,2}\rightarrow +\infty$ \\
In this limit, the following asymptotic form is deduced for the molecule 2 and it turns out to be
\bea
&&\hspace{-0.4cm}v(x,t)=\frac{1}{D}\bigg(k_4^2(4\mu^2-k_3^2)^\frac{1}{2}\cosh(\eta_3+\vphi_3^+)+k_3^2(4\mu^2-k_4^2)^\frac{1}{2}\cosh(\eta_4+\vphi_4^+)\nonumber\\&&\hspace{1.0cm}-2\mu(k_3^2-k_4^2)\bigg),\label{53}\\
&&\hspace{-0.4cm}D=\frac{c_3}{2}(4\mu^2-k_3^2)^\frac{1}{2}(4\mu^2-k_4^2)^\frac{1}{2}\cosh(\eta_3+\eta_4+\vphi_3^++\vphi_4^+)\nonumber\\&&\hspace{0.1cm}+\frac{1}{2c_3}(4\mu^2-k_3^2)^\frac{1}{2}(4\mu^2-k_4^2)^\frac{1}{2}\cosh(\eta_3-\eta_4+\vphi_3^+-\vphi_4^+)\nonumber\\
&&\hspace{0.1cm}-2\mu(4\mu^2-k_3^2)^\frac{1}{2}\cosh(\eta_3+\vphi_3^+)-2\mu(4\mu^2-k_4^2)^\frac{1}{2}\cosh(\eta_4+\vphi_4^+)+c_4,\nonumber
\eea
where the phase terms are given by $\vphi_3^+=\frac{1}{2}\log\frac{(k_1-k_3)^4(k_2-k_3)^4(k_3-k_4)^2(4\mu^2-k_3^2)}{4k_3^4(k_3+k_4)^2(k_1+k_3)^4(k_2+k_3)^4}$, $\vphi_4^+=\frac{1}{2}\log\frac{(k_1-k_4)^4(k_2-k_4)^4(k_3-k_4)^2(4\mu^2-k_4^2)}{4k_4^4(k_3+k_4)^2(k_1+k_4)^4(k_2+k_4)^4}$. From the above asymptotic expressions, we find that the phase terms corresponding to the molecules after and before collision are related as  follows:  
\bes\bea
&&\vphi_1^+=\vphi_1^--\frac{1}{2}\log\frac{(k_1-k_3)^4(k_1-k_4)^4}{(k_1+k_3)^4(k_1+k_4)^4}, \label{54a}\\ &&\vphi_2^+=\vphi_2^--\frac{1}{2}\log\frac{(k_2-k_3)^4(k_2-k_4)^4}{(k_2+k_3)^4(k_2+k_4)^4},\label{54b}\eea\bea &&\vphi_3^+=\vphi_3^-+\frac{1}{2}\log\frac{(k_1-k_3)^4(k_2-k_3)^4}{(k_1+k_3)^4(k_2+k_3)^4},\label{54c}\\ &&\vphi_4^+=\vphi_4^-+\frac{1}{2}\log\frac{(k_1-k_4)^4(k_2-k_4)^4}{(k_1+k_4)^4(k_2+k_4)^4}.\label{54d}
\eea\ees

\begin{figure}[]
	\centering
	\includegraphics[width=0.85\linewidth]{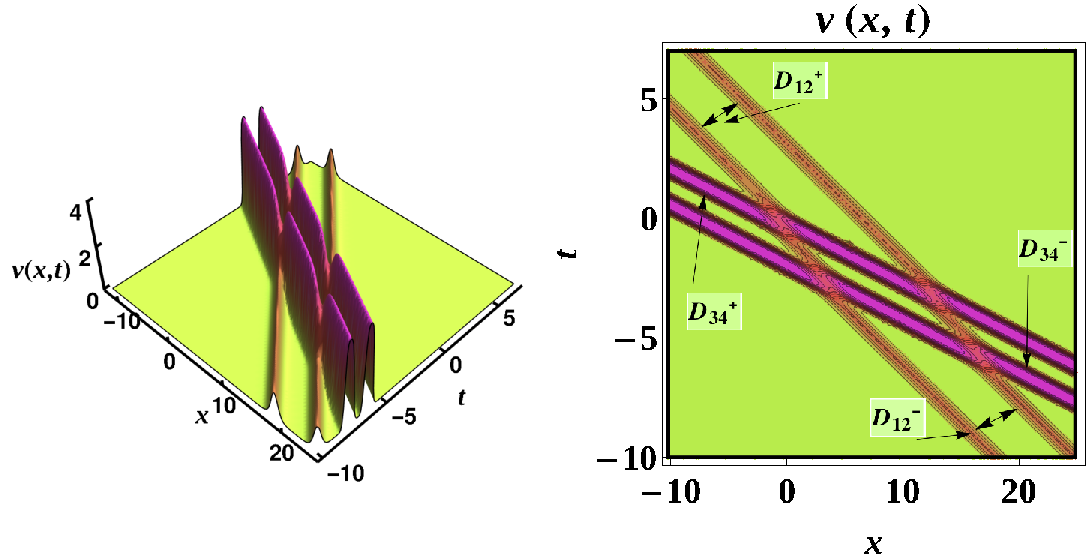}
	\caption{An elastic collision between a dissociated molecule and a synthetic molecule.  }
	\label{fg9}
\end{figure}

\begin{figure}[]
	\centering
	\includegraphics[width=0.85\linewidth]{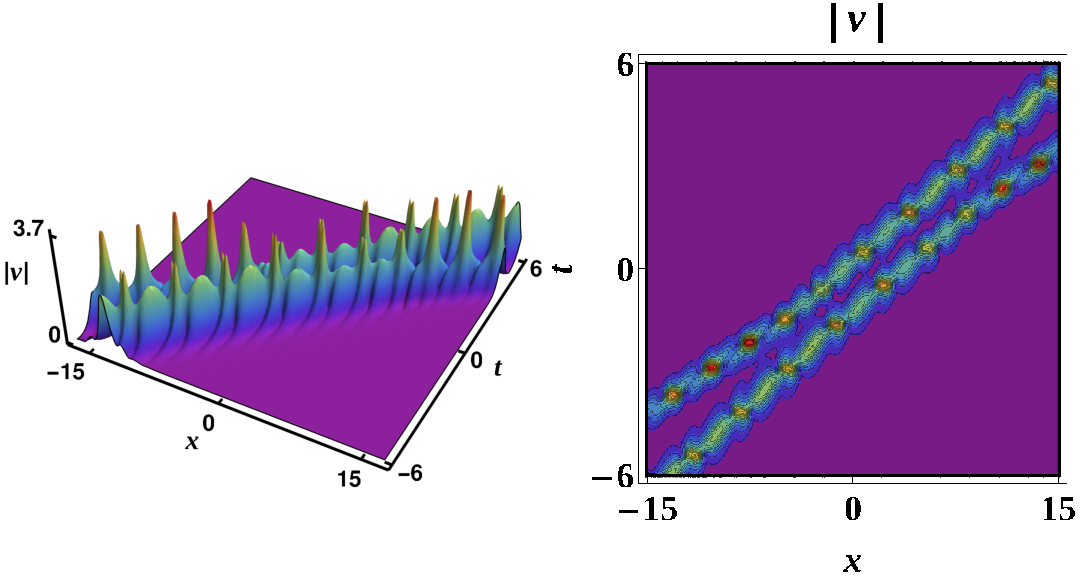}
	\caption{Collision among the two complexitons are demonstrated by considering the wave number resonance condition $k_2=k_1^*$, and $k_4=k_3^*$.}
	\label{fg10}
\end{figure}

From the above analysis, we observe that the two fundamental SMs preserve their structures even after collision with each other. One can easily understand it from the asymptotic expressions (\ref{50})-(\ref{53}) and from Fig. \ref{fg9}, where the collision scenario between a dissociated molecule and a synthetic bi-soliton molecule is depicted for the values $k_1=-2.886$, $k_2=-2.887$, $k_3=-3.9998$, $k_4=-3.9999$, $\eta_1^{(0)}=2$, $\eta_2^{(0)}=0$, and $\eta_{3,4}^{(0)}=0$. This ensures that the soliton molecules always exhibit elastic collision only. However, from Fig. \ref{fg9} we observe that each bi-SM encounters a group phase shift. This  occurs essentially  because the individual soliton atoms in a molecule undergo two consecutive phase shifts with the constituents of another molecule. As a consequence, the variations in bond lengths of each of the molecules occur. The variation in the bond length of the molecule 1 is calculated as  
 \bes\bea
 &&\hspace{-1.0cm}\Delta x_{12}^-~ (D_{12}^-)=\frac{\eta_2^{(0)}}{k_2}+\frac{1}{2k_2}\log\frac{(4\mu^2-k_2^2)(k_1-k_2)^2(k_2-k_3)^4(k_2-k_4)^4}{4k_2^4(k_1+k_2)^2(k_2+k_3)^4(k_2+k_4)^4}\nonumber\\
 &&~~~~~-\frac{1}{2k_1}\log\frac{(4\mu^2-k_1^2)(k_1-k_2)^2(k_1-k_3)^4(k_1-k_4)^4}{4k_1^4(k_1+k_2)^2(k_1+k_3)^4(k_1+k_4)^4}-\frac{\eta_1^{(0)}}{k_1},~~ \label{55a} \\
 &&\hspace{-1.0cm}\Delta x_{12}^+~(D_{12}^+)=\frac{\eta_2^{(0)}}{k_2}+\frac{1}{2k_2}\log\frac{(4\mu^2-k_2^2)(k_1-k_2)^2}{4k_2^4(k_1+k_2)^2}\nonumber\\
 &&~~~~~~~~-\frac{\eta_1^{(0)}}{k_1}-\frac{1}{2k_1}\log\frac{(4\mu^2-k_1^2)(k_1-k_2)^2}{4k_1^4(k_1+k_2)^2}. \label{55b}
 \eea\ees  
The total variation in the bond length of the bi-soliton molecule 1 is calculated as
\bea
\Del \Phi_{12}=\frac{1}{2k_2}\log\frac{(k_2+k_3)^4(k_2+k_4)^4}{(k_2-k_3)^4(k_2-k_4)^4}-\frac{1}{2k_1}\log\frac{(k_1+k_3)^4(k_1+k_4)^4}{(k_1-k_3)^4(k_1-k_4)^4}.
\eea 
Similarly, the variation in the bond length of the molecule 2 is calculated as
\bes\bea
&&\hspace{-1.0cm}\Delta x_{34}^+~ (D_{34}^+)=\frac{\eta_4^{(0)}}{k_4}+\frac{1}{2k_4}\log\frac{(4\mu^2-k_4^2)(k_3-k_4)^2(k_2-k_4)^4(k_1-k_4)^4}{4k_4^4(k_3+k_4)^2(k_2+k_4)^4(k_1+k_4)^4}\nonumber\\
&&~~~~~-\frac{1}{2k_3}\log\frac{(4\mu^2-k_3^2)(k_3-k_4)^2(k_1-k_3)^4(k_2-k_3)^4}{4k_3^4(k_3+k_4)^2(k_1+k_3)^4(k_2+k_3)^4}-\frac{\eta_3^{(0)}}{k_3},~~ \label{57a} \\
&&\hspace{-1.0cm}\Delta x_{34}^-~(D_{34}^-)=\frac{\eta_4^{(0)}}{k_4}+\frac{1}{2k_4}\log\frac{(4\mu^2-k_4^2)(k_3-k_4)^2}{4k_4^4(k_3+k_4)^2}\nonumber\\
&&~~~~~~~~-\frac{\eta_3^{(0)}}{k_3}-\frac{1}{2k_3}\log\frac{(4\mu^2-k_3^2)(k_3-k_4)^2}{4k_3^4(k_3+k_4)^2}. \label{57b}
\eea\ees  
The total variation in the bond length of the bi-soliton molecule 2 is calculated as
\bea
\Del \Phi_{34}=\frac{1}{2k_4}\log\frac{(k_1-k_4)^4(k_2-k_4)^4}{(k_1+k_4)^4(k_2+k_4)^4}-\frac{1}{2k_3}\log\frac{(k_1-k_3)^4(k_2-k_3)^4}{(k_1+k_3)^4(k_2+k_3)^4}.
\eea 
Based on the above calculations, one can say that during the collision process the molecular structures either  get distorted slightly or remain constant depending on the choice of the parameter values.
However, in general, it shows that there is a possibility of occurrence of slight deformation in the molecular structure. By using the parameter values corresponding to Fig. \ref{fg9}, we compute the bond length values. The numerical values corresponding to bond lengths, before and after collision, of soliton molecule 1  are $\Delta x_{12}^-=-0.691545$, and $\Delta x_{12}^+=-0.692116$, respectively. Similarly, for soliton molecule 2, these values are obtained as $\Delta x_{34}^-=0.0865442$, and $\Delta x_{34}^+=0.0861056$, respectively. From these values we infer that only negligible variations occurred in the bond lengths of both the SMs. Therefore, the collision scenario that occurs in between the two bi-soliton molecules is always elastic as it is evident from Fig. \ref{fg9}, where a stable propagation of SMs is observed. We also illustrate the collision among the two complexitons in Fig. \ref{fg10} for $k_1=1+2i$, $k_2=k_1^*$, $k_3=1.3+2.3i$, $k_4=k_3^*$, $\eta_1^{(0)}=0$, $\eta_2^{(0)}=1$, $\eta_3^{(0)}=1+i$, and $\eta_4^{(0)}=1.2+2i$.   
\section{Soliton molecules in FPUT-$\beta$ model: Modified KdV equation}
\begin{figure}[]
 	\centering
 	\includegraphics[width=1.0\linewidth]{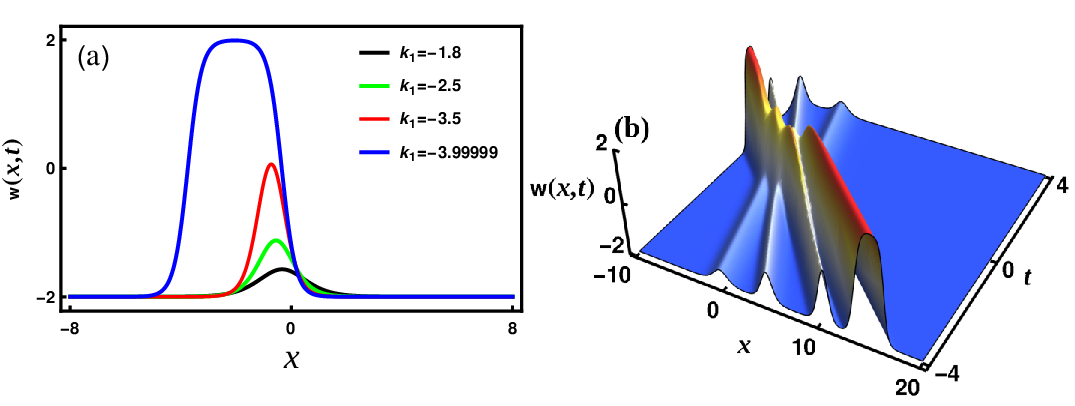}
 	\caption{The various profiles of fundamental soliton of the modified KdV equation (\ref{10}) is displayed in Fig. (a). A flat-top soliton excitation (blue line) arises for  $k_1=-3.99999$. The standard elastic collision among the four solitons are depicted in Fig. (b) for $k_1=-2$, $k_2=-2.8$, $k_3=-3.5$, $k_4=-3.99999$, $\eta_{1,2}^{(0)}=0$, $\eta_{3}^{(0)}=1$, and $\eta_{4}^{(0)}=2$.   }
 	\label{fg11}
 \end{figure}

Now, we demonstrate the existence of soliton molecules in the FPUT-$\beta$ (Eq. (\ref{fput}) with cubic nonlinearity) model based on the soliton solutions of the modified KdV equation (\ref{10}). This is because of the fact that mKdV equation can be deduced asymptotically  from the FPUT model (\ref{fput}) when the cubic ($\beta\neq 0$ and $\alpha=0$) nonlinear interaction  between the  mass points alone is taken into account. To display the existence of soliton molecules, one has to deduce the high-order soliton solutions for the mKdV equation by using the known soliton solutions of the Gardner equation (\ref{1}).  Before proceeding further, first let us discuss about the nature of the fundamental soliton solution in the mKdV Eq. (\ref{10}). We obtain the following fundamental soliton solution of the mKdV equation (\ref{10}) by utilizing the one-soliton solution (\ref{17}) in $w=v+\mu$, $\mu=-\sqrt{\frac{2a_1}{3}}$. The resultant action yields
\bea
w(x,t)=\frac{k_1^2-2\mu^2+\mu\sqrt{4\mu^2-k_1^2}\cosh(\eta_1+\frac{\Del_1}{2})}{-2\mu+\sqrt{4\mu^2-k_1^2}\cosh(\eta_1+\frac{\Del_1}{2})},\label{59}
\eea
where $\eta_1=k_1x+\frac{k_1^3}{4}t+\eta_1^{(0)}$, and $\frac{\Del_1}{2}=\frac{1}{2}\log\frac{4\mu^2-k_1^2}{4k_1^4}$. In a similar way one has to deduce the higher-order soliton solutions for Eq. (\ref{10}) by making use of multi-soliton solutions (Eqs. (\ref{20a})-(\ref{20b}) and Eqs. (\ref{21a})-(\ref{21b})) of the Gardner equation (\ref{1}). An interesting feature of the above one-soliton solution (\ref{59}) is that it admits both sech-type and table-top profiles with constant background, which can be determined from the value of $\mu$ (or $a_1$). We display such profiles associated with the solution (\ref{59}) in Fig. \ref{fg11}(a) for  various values of $k_1$ and $a_1=6$. The figure shows that the soliton profiles persist on the constant background $\mu=-2$. We remark here that the explicit form of the soliton solution (\ref{59}) derived by us has not been reported so far for the  mKdV equation (\ref{10}) but its existence was pointed out in \cite{zimmer}. When $\mu=0$ the solution (\ref{59}) becomes the standard mKdV pulse soliton of complex type. The higher-order solitons of type (\ref{59}) exhibit the standard elastic collision which is demonstrated in Fig. \ref{fg11}(b). This figure illustrates that the four solitons preserve their characters even after collision.    
\begin{figure}[]
	\centering
	\includegraphics[width=0.8\linewidth]{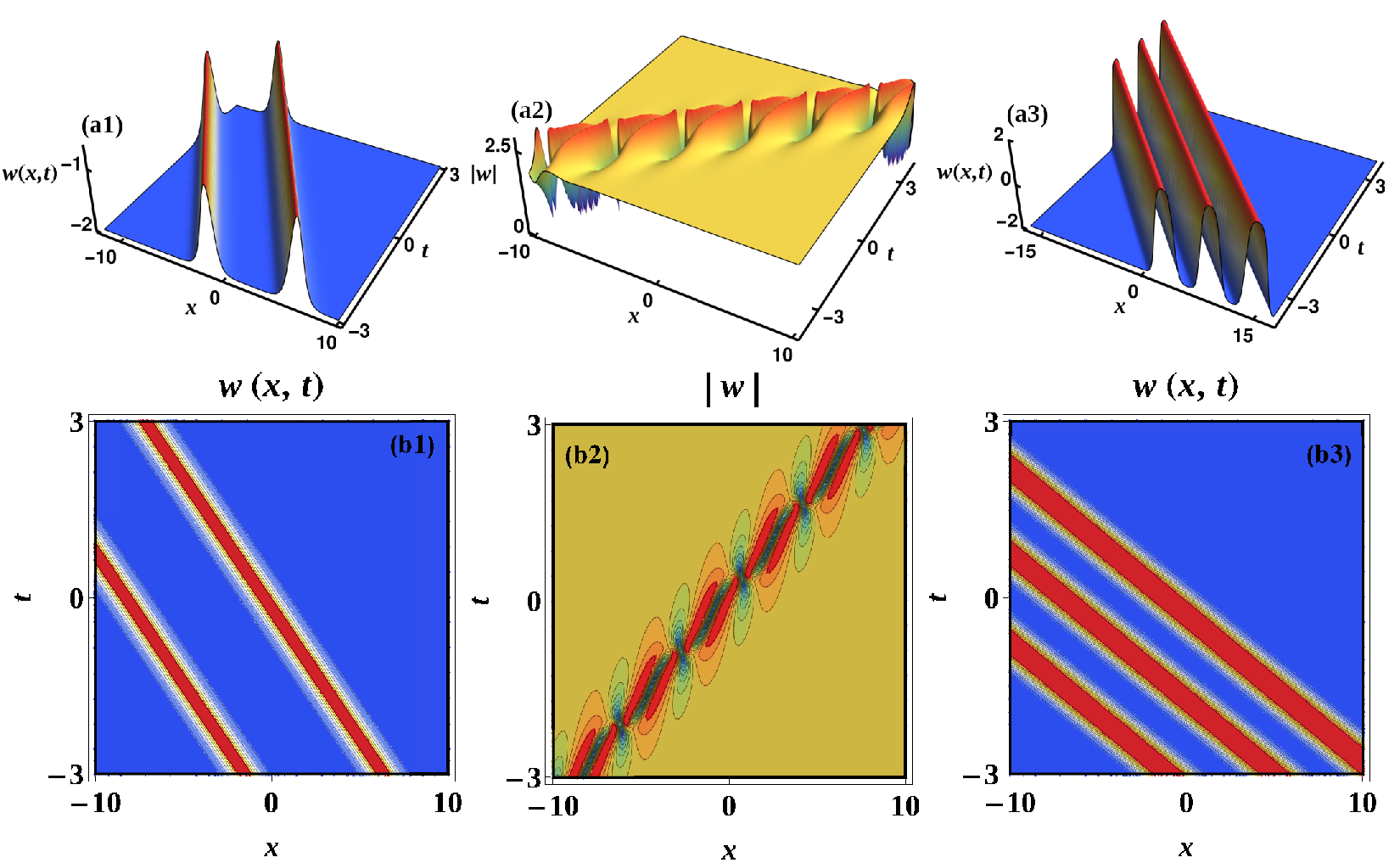}
	\caption{A dissociated sech-type bi-soliton molecule is illustrated in Figs. (a1)-(b1) for $k_1=-2.99981$, $k_2=-2.99982$, $\eta_1^{(0)}=1$, and $\eta_2^{(0)}=0$. In Fig. (a2)-(b2), we displayed the complexiton for $k_2=1+2i=k_1^*$, and  $\eta_2^{(0)}=\eta_1^{(0)*}=0$.   A synthetic tri-soliton molecule is demonstrated in Figs. (a3)-(b3) for $k_1=-3.9997$, $k_2=-3.9998$, $k_3=-3.9999$,  and $\eta_j^{(0)}=0$, $j=1,2,3$. }
	\label{fg12}
\end{figure}

\begin{figure}[]
 	\centering
 	\includegraphics[width=0.8\linewidth]{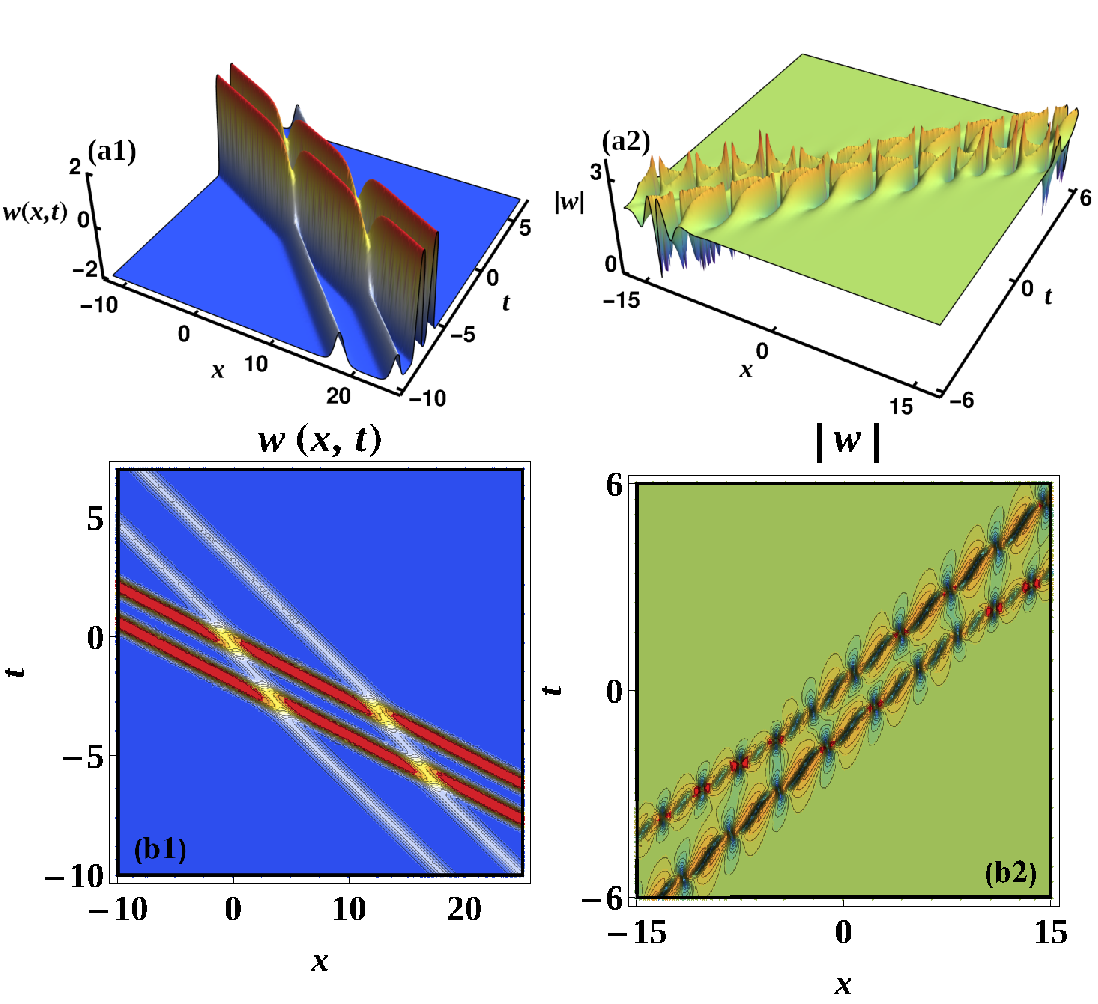}
 	\caption{An elastic collision among the two bi-soliton molecules is illustrated in Fig. (a1)-(b1) for $k_1=2.886$, $k_2=2.887$, $k_3=-3.9998$, $k_4=-3.9999$, $\eta_{1}^{(0)}=1$, $\eta_{2}^{(0)}=0$, $\eta_{3,4}^{(0)}=0$. In Fig. (a2)-(b2), we depicted the collision among two-complexitons. }
 	\label{fg13}
 \end{figure}

As we have pointed out in the Gardner case, to obtain the fundamental SM for the mKdV Eq. (\ref{10}), one has to get the explicit form of the two-soliton solution. Once it is known, to obtain the bi-soliton molecule we impose the resonance condition (\ref{30}) on it. This action yields bi-soliton molecule of either synthetic type or dissociated type. A typical profile of dissociated bi-soliton molecule of Eq. (\ref{10}) is demonstrated as an example in Figs. \ref{fg12}(a1)-(b1). Then, by proceeding in the same way along with the velocity resonance condition (\ref{31}), we have arrived at the tri-soliton molecule of the mKdV equation (\ref{10}). Such a tri-soliton molecule of table-top type is depicted in Figs. \ref{fg12}(a3)-(c3). Then for completeness in Figs. \ref{fg12}(a2)-(c2), we also demonstrate the profile of complexiton in the mKdV case by considering the complex wave numbers $k_2=k_1^*$. Further, to verify the stability and collision properties associated with the obtained SMs, we consider the interaction among the two-SMs as it is demonstrated in Figs. \ref{fg13}(a1)-(b1). From these figures we observe that the shapes of the molecules remain unchanged except for the group phase shifts. Due to these group phase shifts, the bond-lengths of the both the molecules may slightly vary (one can verify this by calculating the spatial separation between the soliton atoms as we have done earlier in the case of Gardner Eq. (\ref{1})). However, their entire molecular structures continue to be preserved throughout the collision process as it is true from Figs. \ref{fg13}(a1)-(b1). Therefore the bi-SMs, in the mKdV equation case, are stable against soliton collision and propagate without any deformation, except for the changes in the bond lengths. In addition, we also display the two-complexiton profiles in Figs. \ref{fg13}(a2)-(b2) for $k_1=1+2i$, $k_2=k_1^*$, $k_3=1.3+2.3i$, $k_4=k_3^*$, $\eta_{1,3}^{(0)}=0$, $\eta_2^{(0)}=1+i$, and $\eta_4^{(0)}=1$.   
 \begin{figure}[]
	\centering
	\includegraphics[width=0.7\linewidth]{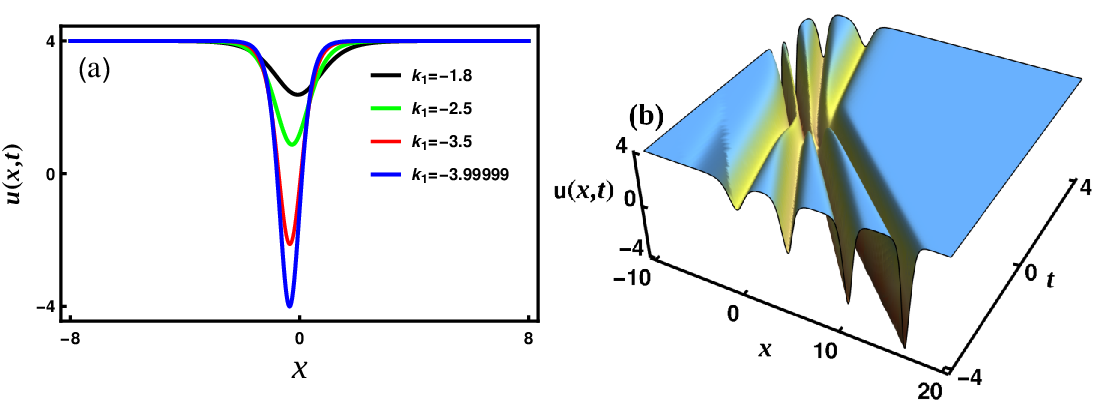}
	\caption{The various profiles corresponding to the fundamental dark soliton of the KdV equation is illustrated in Fig. (a). In Fig. (b), we demonstrate the standard elastic collision of the four dark-solitons. To bring out this collision picture we set the parameter values as  $k_1=-2$, $k_2=-2.8$, $k_3=-3.5$, $k_4=-3.99999$, $\eta_{1,2}^{(0)}=0$, $\eta_{3}^{(0)}=1$, and $\eta_{4}^{(0)}=2$. }
	\label{fg14}
\end{figure}
\section{Soliton molecules in FPUT-$\alpha$ model: KdV equation}
We find that the well known KdV equation (\ref{11}) arises in the continuous limit of the FPUT lattice model (\ref{fput})  while considering the quadratic ($\alpha\neq 0$ and $\beta=0$) nonlinear interaction. To exhibit the excitation of soliton molecules in this case, we derive the soliton solutions for the KdV equation with the help of the soliton solutions of the mKdV Eq. (\ref{10}) through the Miura transformation (\ref{13}). By doing so, we deduce the following fundamental soliton solution of the KdV equation (\ref{11}) by substituting the solution (\ref{59}) in the transformation (\ref{13}). This action yields the form of the one-soliton solution of the KdV equation as 
\bea
&&u(x,t)=\frac{1}{D^2}(2\mu^2-k_1^2)\bigg[2\mu^2-k_1^2-2\mu\sqrt{4\mu^2-k_1^2}\cosh(\eta_1+\frac{\Del_1}{2})\bigg] \label{60}\\
&&\hspace{1.3cm}+\sqrt{4\mu^2-k_1^2}\bigg[\mu^2\sqrt{4\mu^2-k_1^2}\cosh^2(\eta_1+\frac{\Del_1}{2})-k_1^3\sinh(\eta_1+\frac{\Del_1}{2})\bigg],\nonumber\\
&&D=\bigg[-2\mu+\sqrt{4\mu^2-k_1^2}\cosh(\eta_1+\frac{\Del_1}{2})\bigg],\nonumber
\eea
where $\eta_1=k_1x+\frac{k_1^3}{4}t+\eta_1^{(0)}$, and $\Del_1=\log\frac{4\mu^2-k_1^2}{4k_1^4}$.  We remark that the above resultant soliton solution (\ref{60}) is different from the standard ``$\sech^2$" KdV soliton solution since it takes over the constant boundary condition from the non-vanishing boundary condition of the solution of the mKdV equation. An interesting aspect of the solution (\ref{60}) is that it admits dark soliton profiles on the constant background as it was widely discussed in nonlinear optics \cite{kivshar}. The various profiles of the dark-soliton is drawn in Fig. \ref{fg14}(a) by considering different values of the wave number $k_1$. 
This special kind of dark soliton solution for the KdV equation has not been reported elsewhere in the literature except in Ref. \cite{zimmer}, where the existence of dark soliton in the KdV equation is just indicated. We note that by following the same methodology as discussed earlier, one has to deduce the higher-order soliton solutions of the KdV equation by making use of the higher-order solutions of the modified KdV equation. We have omitted these details for brevity. Then, we graphically demonstrate the collision of multi-dark solitons in Fig. \ref{fg14}(b). As it is evident from this figure, the amplitudes of the four dark solitons remain unchanged throughout the collision process thereby ensuring the elastic nature of the collision.       

\begin{figure}[]
	\centering
	\includegraphics[width=0.85\linewidth]{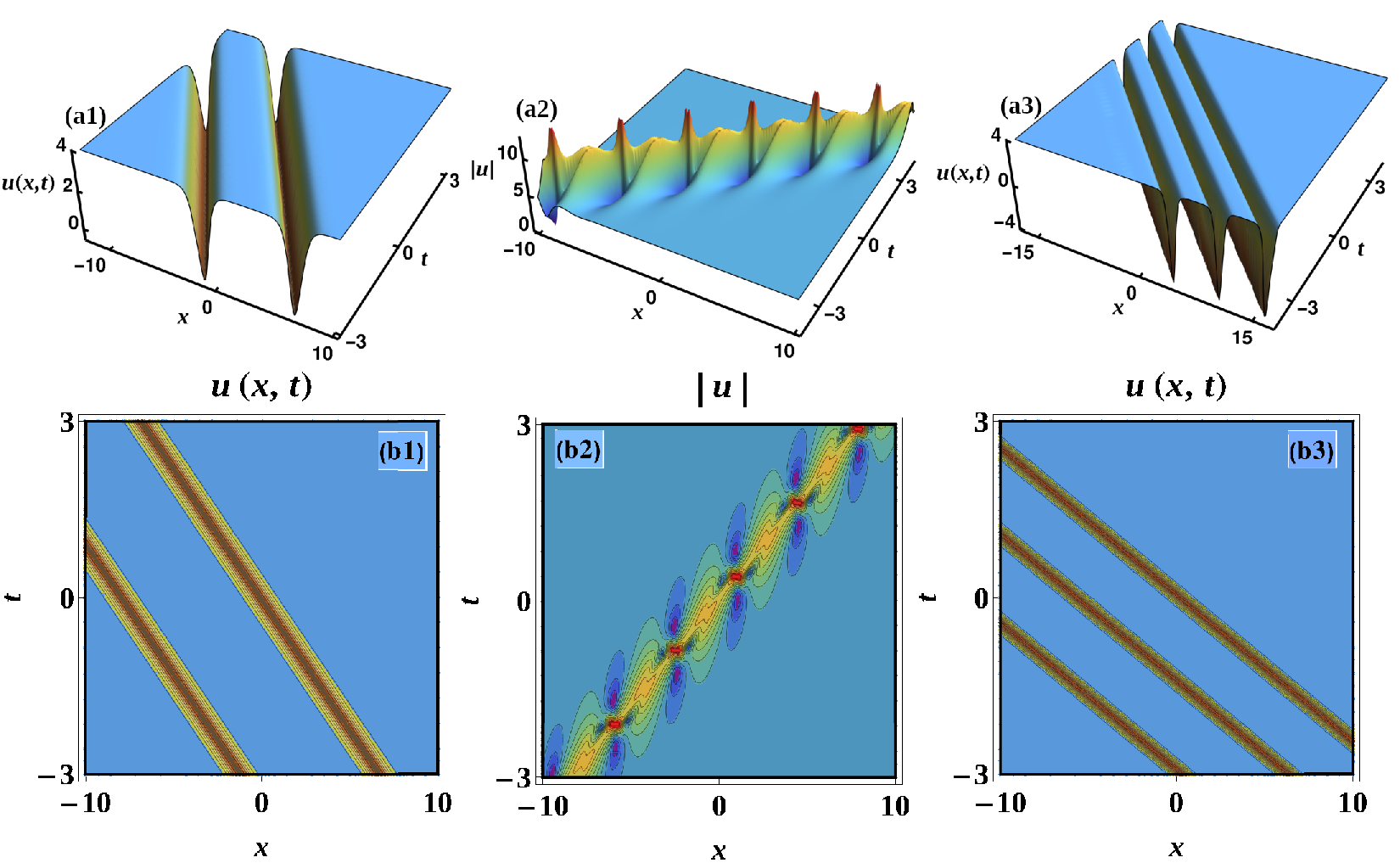}
	\caption{ In Figs. (a1)-(b1) and (a3)-(b3) we have displayed the profiles of bi- and tri-dark-soliton molecules, respectively, of the KdV equation. A complexiton profile is demonstrated in Fig. (a2)-(b2).   }
	\label{fg15}
\end{figure}
To bring out the existence of bi- and tri-dark soliton molecules for the KdV case, we have computed the two- and three-soliton solutions  through the transformation (\ref{13}) using the two- and three-soliton solutions of the mKdV equation (\ref{10}), respectively.  By doing so, along with the velocity resonance conditions (\ref{30}) and (\ref{31}), we arrive at the forms of bi- and tri-soliton molecules of the KdV equation. We have depicted their profiles in Figs. \ref{fg15}(a1)-(b1) and   \ref{fg15}(a3)-(b3), respectively. To get the bi-SM profile (Figs. \ref{fg15}(a1)-(b1)), we set the parameter values as $k_1=-2.99981$, $k_2=-2.99982$, $\eta_1^{(0)}=1$, and $\eta_2^{(0)}=0$. Then to draw the profile of tri-SM of the KdV Eq. (\ref{11}) we set the values as $k_1=-3.9997$, $k_2=-3.9998$, $k_3=-3.9999$,  and $\eta_j^{(0)}=0$, $j=1,2,3$. Then a complexiton profile is brought out in   Figs. \ref{fg15}(a2)-(b2), for $k_2=1+2i=k_1^*$, and  $\eta_2^{(0)}=\eta_1^{(0)*}=0$.   
\begin{figure}[]
	\centering
	\includegraphics[width=0.8\linewidth]{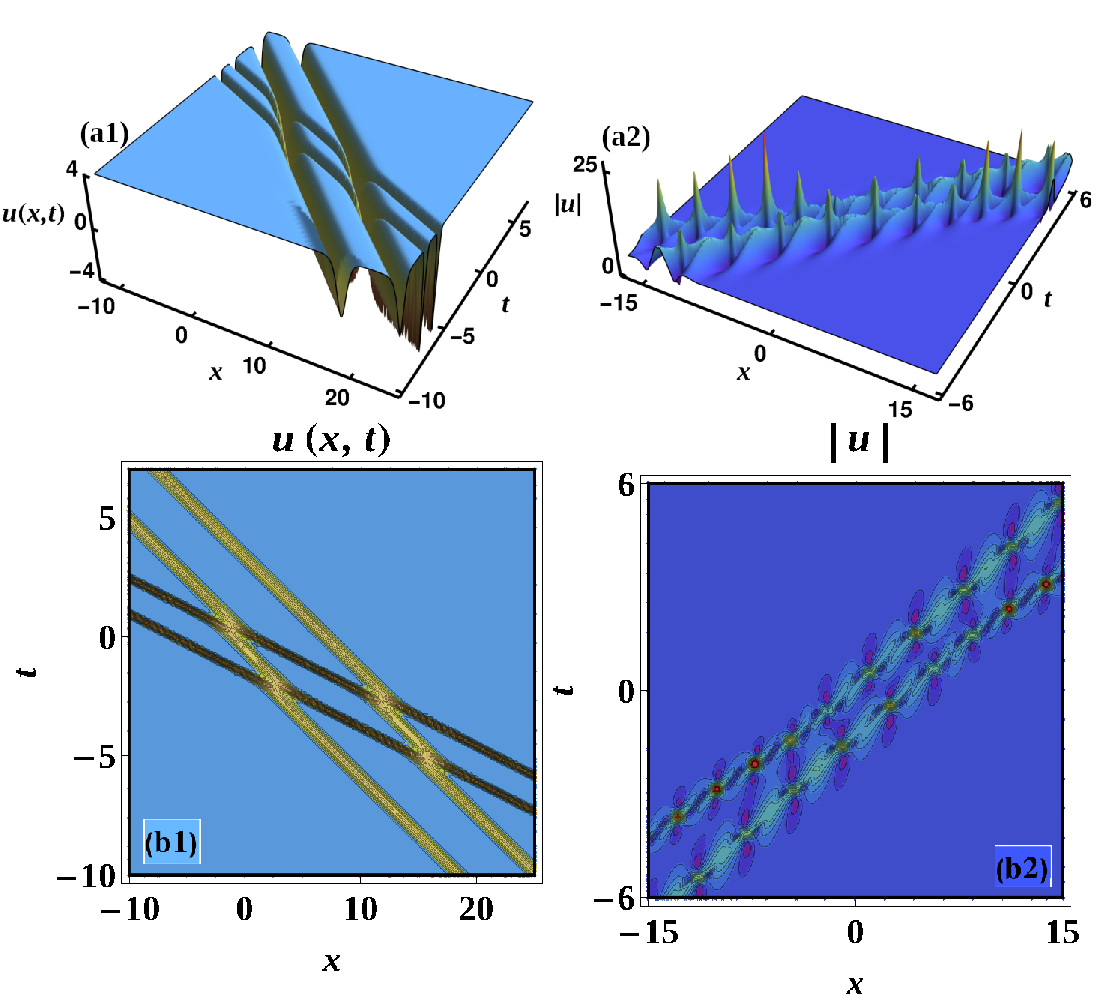}
	\caption{A shape preserving collision among the two bi-dark soliton molecules of the KdV equation is illustrated in Figs. (a1)-(b1) for $k_1=2.886$, $k_2=2.887$, $k_3=-3.9998$, $k_4=-3.9999$, $\eta_{1}^{(0)}=1$, $\eta_{2}^{(0)}=0$, $\eta_{3,4}^{(0)}=0$. In Figs. (a2)-(b2), we have depicted the collision among two-complexiton solution of the KdV equation.}
	\label{fg16}
\end{figure}

Next, to verify the stability associated with the obtained SMs, we allow two bi-dark SMs to interact with themselves. Such a collision scenario among the two-soliton molecules is demonstrated in Figs. \ref{fg16}(a1)-(b1). From these figures, we understand that the structures of the both molecules remain the same except for the variations in the group phase shifts. Due to the group phase shifts, the bond-lengths slightly vary. However, the slight variation in bond-lengths does not completely deform the shapes of the entire molecular structures. After the collision process, they continue to propagate in a stable manner without any further distortion. To confirm this, one has to calculate the relative separation distances between the soliton atoms which can be done by following the earlier procedure. Therefore, the bi-dark soliton molecules in the KdV equation are stable against collision. In addition, we also display the two-complexiton profiles in Fig. \ref{fg16}(a2)-(b2) for $k_1=1+2i$, $k_2=k_1^*$, $k_3=1.3+2.3i$, $k_4=k_3^*$, $\eta_{1,3}^{(0)}=0$, $\eta_2^{(0)}=1+i$, and $\eta_4^{(0)}=1$.      
\begin{figure}[]
	\centering
	\includegraphics[width=0.91\linewidth]{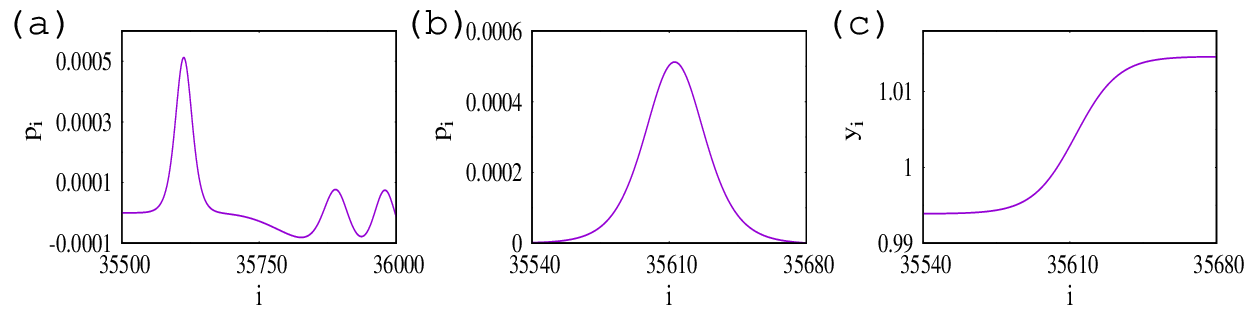}
	\caption{In Fig. (a), we demonstrate the emergence of a smooth soliton from the wave group whereas in Fig. (b) we display the fully isolated single smooth soliton. In Fig. (c) we demonstrate the kink-soliton by plotting $y_i$. These figures are drawn for $m=1$, $a=1$, $c=1$, $N=400000$, $\alpha=2.4$, $\beta=-0.4$, $\mu=-2$, $k=1$, $u_0=1$ and $k_1=-0.12$. } 
	\label{fg17}
\end{figure}
\begin{figure}[]
	\centering
	\includegraphics[width=0.8\linewidth]{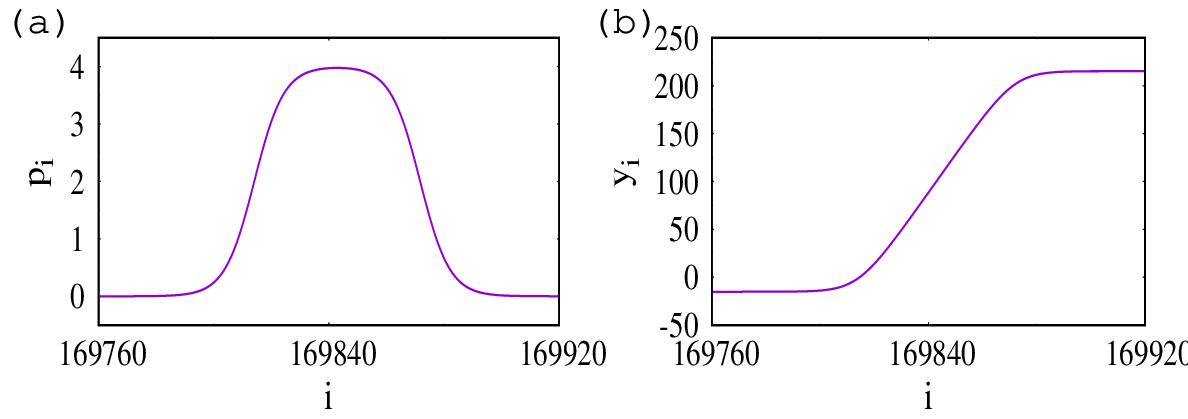}
	\caption{In Fig. (a), we demonstrate the fully isolated flat-top soliton from the wave train whereas in Fig. (b) draw a kink-soliton by plotting the corresponding solution space $y_i$. We draw these figures by choosing the parameter values as $a=0.01$, $m=1$, $N=400000$, $\alpha=0.0024$, $\beta=-0.0004$, $k=1$, $k_1=-3.9999$, $c=0.01$, $\mu=-2$, and $u_0=1$.}
	\label{fg18}
\end{figure}
\section{Numerical simulation of the FPUT lattice: Excitation of solitons and soliton molecule}
In this section, we verify the analytical results obtained for the Gardner equation in the continuous limit by  simulating the FPUT chain with quadratic and cubic nonlinear interactions numerically by adopting the fourth-order Runge-Kutta algorithm. In particular, we intend to confirm the existence of fundamental smooth sech-type and flat-top solitons, their collision properties and flat-top soliton molecular structure in the discrete domain. The following soliton solution, for the displacement ($y_i$) and momentum ($\displaystyle{p_i=m \dot{y}_i}$, $i=1,2,...N$), in the discrete domain is computed by using the soliton solution (\ref{18}) of the Gardner equation in the continuous limit, 
\begin{subequations}
\begin{eqnarray}
&&\hspace{-1.0cm}y_i=\frac{1}{\sqrt{6}a}\tanh^{-1}\bigg(\frac{\gamma}{k_1}\tanh\big(\sqrt{6}k_1(-ia+c(1+a^2k_1^2)t)+\frac{\Delta_1}{4}\big)\bigg)+\frac{u_0}{a},\label {61a}~\\
&&\hspace{-1.0cm}p_i=m\dot{y}_i=\frac{mc(1+a^2k_1^2)\gamma \sech^2\big(\sqrt{6}k_1(-ia+c(1+a^2k_1^2)t)+\frac{\Delta_1}{4}\big)}{a\bigg(1-\frac{\gamma^2}{k_1^2}\tanh^2(\sqrt{6}k_1(-ia+c(1+a^2k_1^2)t)+\frac{\Delta_1}{4})\bigg)},\label {61b}
\end{eqnarray}
\end{subequations}
where, $i=1,2,...,N$, $\gamma=2\mu+\sqrt{4\mu^2-k_1^2}$, $\frac{\Delta_{1}}{4}=\frac{1}{4}\ln\frac{4\mu^2-k_1^2}{4k_1^4}$, and $u_0$ is an integration constant. The above soliton solution is valid by neglecting the higher-order derivatives. This can be true when $6k_1^2<<1$. The solution $y_i$ and momentum $p_i$ are kink and table-top/sech-type soliton, respectively. The soliton velocity is calculated from Eqs. (\ref{61a})-(\ref{61b}) as $v_s=c(1+a^2k_1^2)$. To calculate the soliton energy, we consider the Hamiltonian of the FPUT lattice
\begin{eqnarray}
H=\sum_i^{N} \frac{p_i^2}{2}+V(y_{i+1}-y_{i}), 
\end{eqnarray}
where $\displaystyle{V(y_{i+1}-y_i)=\frac{k}{2}(y_{i+1}-y_i)^2+\frac{\alpha }{3}(y_{i+1}-y_i)^3+\frac{\beta }{4}(y_{i+1}-y_i)^4}$. In the continuum or long wavelength limit $a\rightarrow 0$, the potential energy is close to the kinetic energy so that the Hamiltonian can be expressed as $H=2\sum_i^N p_i^2/2$. Therefore, the soliton energy at time $t=0$ is obtained by substituting the expression of $p_i$ in the latter energy expression. This results in the soliton energy 
\begin{eqnarray}
E=\sum_i^N \frac{c^2(1+a^2k_1^2)^2\gamma^2\sech^4(k_1\sqrt{6}(-ia+\frac{\Delta_1}{4}))}{a^2\big(1-\frac{\gamma^2}{k_1^2}\tanh^2(k_1\sqrt{6}(-ia+\frac{\Delta_1}{4}))\big)^2}.
\end{eqnarray}
We arrive at the soliton energy, by replacing the above summation by integral, as $E=-\frac{c^2k_1}{a^2\sqrt{6}}$. From the latter energy expression, one can get the relation between soliton velocity and energy as $\displaystyle{v_s=c(1+\frac{6a^6E^2}{c^4})}$. 

For the numerical simulation, the initial conditions are considered from Eqs. (\ref{61a}) and (\ref{61b}). Then, we consider the soliton excitation under free boundary condition so that the soliton can be observed for a long time without any influence of boundary. We begin with exciting a smooth sech-type soliton by fixing the parameter values, which are given in the figure caption of Fig. \ref{fg17}, in Eqs. (\ref{61a})-(\ref{61b}). The initial profile is set at the right end of the lattice around $i=390000$. On simulating the FPUT chain, we observe that at first small amplitude wave trains start to emerge and they propagate from right to left direction. That is, from $i=400000$ to $i=1$. We call such small waves as low amplitude/energy excitations. After a long time, one of the small waves starts to move faster and  separates from the wave group. This isolated wave continues to propagates forward in the right to the left direction and becomes a single smooth sech-type solitary wave. The emergence of this solitary wave is captured at time $t\approx 3.1\times 10^5$ seconds around the lattice point $i=35610$ and is depicted in momentum space in Figs. \ref{fg17}(a) and \ref{fg17}(b). We note here that if the wave number value is low (or the initial solitary wave solution with low energy), one has to wait for sufficiently long time to capture the isolated solitary waves from the wave tails. The calculated values of velocity ($v\approx 1.0144$) and energy ($E\approx 0.04898$) confirm that the displayed smooth sech-type solitary wave in Fig. \ref{fg17}(b) could be a soliton \cite{zhao}. In Fig. \ref{fg17}(c), the corresponding kink soliton is depicted by plotting the solution space ($y_i$). Then, to show the existence of flat-top soliton excitation in the FPUT lattice, we again fix the parameter values, which are  specified in the figure caption of Fig. \ref{fg18}, in Eqs. (\ref{61a}) and (\ref{61b}), and set the initial profile near the right end of the lattice at $i=290000$. We observe that flat-top structured soliton gets isolated around $i=169840$ lattice points after the time $t=1.2\times 10^5$ seconds from the wave trains made of several small amplitude solitary waves. However, the amplitude and velocity of the flat-top soliton is higher than the solitary wave group so it travels faster and subsequently it becomes separated from the tail. Such flat-top excitation is displayed in Fig. \ref{fg18}(a). In Fig. \ref{fg18}(b), we draw a kink soliton profile of the solution $y_i$. We wish to remark here that to observe a flat-top soliton one has to essentially restrict the value of lattice constant $a$ as less than $0.1$, otherwise one has to wait for a longer time and need highly sophisticated computational facility to capture the flat-top soliton in the FPUT lattice.  

\begin{figure}[]
	\centering
	\includegraphics[width=0.7\linewidth]{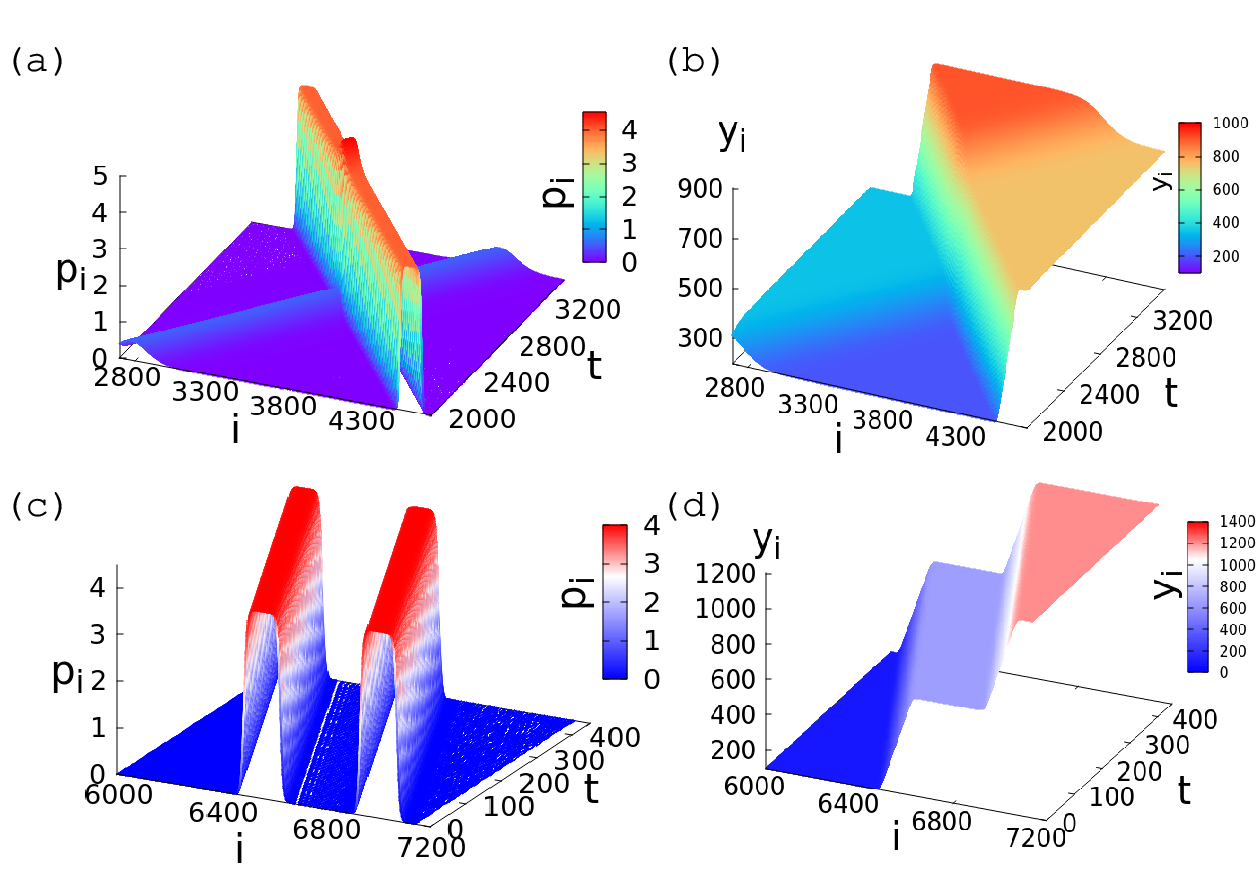}
	\caption{Fig. (a) demonstrates the elastic scattering dynamics between a smooth soliton and a flat-top soliton in momentum space whereas Fig. (b) displays the two kink-soliton scattering in solution space. In Fig. (c) we illustrate the existence of flat-top molecular structure in the FPUT chain by plotting $p_i$ and the corresponding solution space is illustrated in Fig. (d), where two kink-solitons propagate in parallel. }
	\label{fg19}
\end{figure}

Next, we wish to validate the scattering dynamics of the Gardner solitons in the discrete domain. As we have illustrated in Section 3.2, in the discrete case too, the Gardner solitons undergo an elastic collision. To verify this, we again simulate the FPUT lattice with the parameter values $a=0.01$, $m=1$, $c=0.01$, $\alpha=0.0024$, $\beta=-0.0004$, $k=1$, $k_1=0.3$, $k_2=-1$, $\mu=-2$, and $u_0=1$. The latter parametric choice allows us to visualize the elastic head-on scattering dynamics between the smooth and flat-top solitons as illustrated in Fig. \ref{fg19}(a). We prepared the initial conditions of these solitons with the energy ratio of $E_s/E_f=-0.3$, where $E_s$ and $E_f$ are the energies of smooth and flat-top solitons, respectively. From Fig. \ref{fg19}(a), we observe that each of the solitons are well separated initially and the smooth one moves towards the right while the flat-top soliton travels towards the left. After the collision process they pass through each other without changing their shapes and energies. This non-energy exchanging nature of the collision process again confirms that these two solitary waves are solitons. In Fig. \ref{fg19}(b), we have observed the two kink soliton scattering while plotting the corresponding solution $y_i$. Next, we demonstrate the excitation of  synthetic type bi-soliton molecule structure, having two flat-top structured soliton atoms, in Fig. \ref{fg19}(c) by plotting the expression $p_i$. However, the corresponding $y_i$ shows that the two kink-solitons propagate in parallel as it is displayed in Fig. \ref{fg19}(d). To bring out these molecular structure, we choose the wave number values as $k_1=-1$, and $k_2=-0.99$, which obey the velocity resonance condition (\ref{30}). The other parameter values are chosen as $a=0.01$, $\alpha=0.0024$, $\beta=-0.0004$, $k=1$, $c=0.01$, $\mu=-2$, and $u_0=1$. Figs. \ref{fg19}(c) and \ref{fg19}(d) show that the bi-soliton molecule propagates without any distortion over a longer period of time thereby confirming its stable nature. We wish to note that in a similar way one can numerically observe other soliton structures derived in the Gardner equation as well as all the soliton structures that can be deduced analytically for the KdV and mKdV equations by simulating the corresponding FPUT chains.    

\section{Conclusion}
In summary, we reported the existence of an interesting class of soliton molecules in the continuous limit of the FPUT lattice model with quadratic and cubic nonlinear interactions. We have demonstrated their existence based on the multi-soliton solutions of the Gardner equation. To derive these soliton solutions, we have adopted the well known Hirota bilinear method. We have shown the formation of bi-, tri- and quad-soliton molecules based on the velocity resonance mechanism and classified them as dissociated molecules or synthetic molecules, based on their bond-lengths. A more general $N$-soliton molecule can be formed by imposing more restrictions on the velocities of the $N$-solitons. It is found that the bi-, tri- and quad-soliton molecules possess the same properties, for example they do not show oscillations in the coalescence region. However for a certain choice of the  free parameters we have observed the formation of a mega table-top soliton in the overlapping region. It is possible only in the higher-order soliton molecule cases. To confirm the stability properties associated with the obtained soliton molecules, we allow them to interact with the fundamental (or multi) soliton(s). Our asymptotic analysis shows that they are quite stable against soliton collisions, though their bond-lengths can slightly deviate from their initial equilibrium distances. We have also considered the interaction among the two bi-soliton molecules. The results again confirmed that they always undergo elastic collision only. The shape preserving nature of soliton molecules again supports the famous phenomenon of Fermi-Pasta-Ulam-Tsingou recurrence \cite{kruskal,berman}. Besides the above, we also demonstrated the excitation of soliton molecules in the other two sub-cases, namely the FPUT lattice (\ref{fput}) with cubic nonlinear interaction and the FPUT lattice (\ref{fput}) with quadratic nonlinear interaction. To exhibit these possibilities we have deduced the soliton solutions for the mKdV and KdV equations, respectively. The resultant soliton solutions yield an interesting class of soliton molecules for these cases. For instance, table-top soliton molecules with constant background have been observed in the mKdV case whereas for the KdV case the dark-soliton molecules with non-vanishing background have been observed. We find that these molecular structures and their corresponding analytical forms have not been reported earlier in the literature.  In order to validate our results, especially for the Gardner equation, we have numerically simulated the corresponding FPUT chain and verified the various soliton structures that were predicted analytically in the continuous limit. We believe that the results obtained in the present study, in general, can be extended to other integrable and non-integrable systems with applications in Bose-Einstein condensates, nonlinear optics, fluid dynamics, plasma physics and other closely related fields.     

\section*{Declaration of competing interest }
The authors declare that they have no conflict of interest. 
\section*{Acknowledgement}
The works of M. Kirane, and S. Stalin, are supported by Khalifa University of Science and Technology, Abu-Dhabi, UAE, under the Project Grant No. 8474000355. M. Lakshmanan wishes to thank the Department of Science and Technology, INDIA for the award of a DST-SERB National Science Chair under Grant No. NSC/2020/00029 in which R. Arun is supported by a Research Associateship. 

\appendix
\section*{Appendix A. Derivation of the Gardner equation from  the FPUT lattice (\ref{fput})}
To derive the Gardner equation in the continuous limit of the lattice equation (\ref{fput}) one has to consider the combined nonlinear forces, $f(y)=y+\alpha y^2+\beta y^3$, where $\alpha$ and $\beta$  are the strengths of quadratic and cubic nonlinear interactions between the mass points \cite{lakshmanan}. By setting the lattice parameter $a\rightarrow 0$ one can  expand the displacement $y_i$ in terms of Taylor series and the resultant action yields, 
\bea
y_{i\pm 1}(t)= y(x,t)\pm a\frac{\partial y}{\partial x}+\frac{a^2}{2!}\frac{\partial^2 y}{\partial x^2}\pm \frac{a^3}{3!}\frac{\partial^3 y}{\partial x^3}+ \frac{a^4}{4!}\frac{\partial^4 y}{\partial x^4}+... ~. 
\eea
Substituting the above expansion in Eq. (\ref{fput})  with the above mentioned form of force and retaining the series terms upto $O(a^4)$, we obtain the following equation of motion for the continuous case,
\bea
\frac{1}{c^2}\frac{\partial ^2y}{\partial t^2}=\frac{\partial ^2y}{\partial x^2}\bigg[1+2\al a\frac{\partial  y}{\partial x}+3\ba a^2\bigg(\frac{\partial y}{\partial x}\bigg)^2\bigg]+\frac{a^2}{12}\frac{\partial ^4y}{\partial x^4},\label{6}
\eea 
where $c^2=a^2/m$.
Considering the unidirectional propagation of right moving waves and making a change of variables
$\xi=x-ct$, $\tau=a^2ct$, $y=u/a$, we get the following equation from Eq. (\ref{6}), 
\bea
u_{\xi\tau}-2a^2u_{\tau\tau}+\al'u_{\xi}u_{\xi\xi}+\frac{3}{2}\ba'u_{\xi}^2u_{\xi\xi}+\frac{1}{24}u_{\xi\xi\xi\xi}=0,
\eea
where $\alpha'=\alpha/a^2$, $\ba'=\ba/a^2$. By setting  $v=u_{\xi}$ and in the continuous limit $a\rightarrow 0$, we obtain the following form of the Gardner equation (\ref{7}) on replacing $\tau$ and $\xi$ by the standard notation $t=4\sqrt{24}\tau$, and $x=-\sqrt{24}\xi$:
\bea
4v_t-\frac{3}{2}\ba'v^2v_x-\alpha'vv_x-v_{xxx}=0. \label{7}
\eea
For convenience the above equation has been rewritten as the standard form of the Gardner Eq. (\ref{1}). To obtain the Eq. (\ref{1}), we have chosen the constants as $\al'=-12\mu$ and $\ba'=-4$ in Eq. (\ref{7}). Note that here $a_1$ is a real constant. The Gardner equation (\ref{1}) also arises from the mKdV equation, 
\begin{equation}
	4w_t+6w^2w_x-w_{xxx}=0. \label{10}
\end{equation}
To see this clearly, we note that Eq. (\ref{10}) does not possess the Galilean symmetry 
\begin{equation}
	x=x'+a_1t',~~t=t',~~w(x,t)=v(x',t')-\sqrt{\frac{2a_1}{3}}, \label{galilean2}
\end{equation} 
but the above transformation (\ref{galilean2}) yields the Gardner equation (\ref{1}) \cite{zimmer} from the mKdV equation (\ref{10}). On the other hand, it is interesting to note that KdV equation, 
\begin{equation}
	4u_t+6uu_x-u_{xxx}=0,\label{11}
\end{equation}
is invariant under the Galilean transformation,
\begin{equation}
	x=x'+a_1t',~~t=t',~~u(x,t)=u(x',t')+\frac{2}{3}a_1. \label{galilean1}
\end{equation} 

We note here that one can compute the soliton solutions, with non-vanishing boundary condition, of the mKdV equation (\ref{10}) by knowing the soliton solutions of the Gardner equation (\ref{1}) through the relation (\ref{galilean2}) ($w=v-\sqrt{\frac{2a_1}{3}}$) given above. From the relation (\ref{galilean2}), one can understand that the presence of the term, $-\sqrt{\frac{2a_1}{3}}$, provides the non-vanishing background to the final soliton solutions of the mKdV equation. Then with the help of the famous Miura transformation \cite{miura,da-jun-zhang}, which connects the solutions of KdV and mKdV equations, 
\begin{equation}
	u(x,t)=w^2+w_x, \label{13}
\end{equation}
one is able to compute the soliton solutions of the KdV equation (\ref{11}) by using the solutions of the mKdV equation (\ref{10}). Very interestingly, the resultant soliton solutions of the KdV equation inherently follow the boundary condition that $u$ tends to constant as $x\rightarrow \pm \infty$. 
\section*{Appendix B. Three-soliton solution}
The three-soliton solution of the Gardner Eq. (\ref{1}) is  obtaned  as
\bes
\bea
	&&\hspace{-0.5cm}v=\frac{1}{F}\big(e^{\eta_1}+e^{\eta_2}+e^{\eta_3}+e^{\eta_1+\eta_2+\chi_1}+e^{\eta_1+\eta_3+\chi_2}+e^{\eta_2+\eta_3+\chi_3}+e^{2\eta_1+\eta_2+\chi_{11}}\nonumber\\
	&&+e^{\eta_1+2\eta_2+\chi_{12}}+e^{2\eta_1+\eta_3+\chi_{13}}+e^{\eta_1+2\eta_3+\chi_{14}}+e^{2\eta_2+\eta_3+\chi_{15}}\nonumber\\
	&&+e^{\eta_2+2\eta_3+\chi_{16}}+e^{\eta_1+\eta_2+\eta_3+\chi_{17}}+e^{2\eta_1+\eta_2+\eta_3+\chi_{41}}+e^{\eta_1+2\eta_2+\eta_3+\chi_{42}}\nonumber\\
	&&+e^{\eta_1+\eta_2+2\eta_3+\chi_{43}}+e^{2(\eta_1+\eta_2)+\eta_3+\gamma_{1}}+e^{2(\eta_1+\eta_3)+\eta_2+\gamma_{2}}\nonumber\\
	&&+e^{2(\eta_2+\eta_3)+\eta_1+\gamma_{3}}\big),\label{21a}\\
	&&\hspace{-0.5cm}F=1+e^{\eta_1+\delta_1}+e^{\eta_2+\delta_{2}}+e^{\eta_3+\delta_{3}}+e^{2\eta_1+\Delta_1}+e^{2\eta_2+\Delta_{2}}+e^{2\eta_3+\Delta_3}+e^{\eta_2+\eta_1+\delta_{11}}\nonumber\\
	&&+e^{\eta_1+\eta_3+\delta_{12}}+e^{\eta_2+\eta_3+\delta_{13}}+e^{2\eta_1+\eta_2+\delta_{41}}+e^{\eta_1+2\eta_2+\delta_{42}}+e^{2\eta_1+\eta_3+\delta_{43}}\nonumber\\
	&&+e^{\eta_1+2\eta_3+\delta_{44}}+e^{2\eta_2+\eta_3+\delta_{45}}+e^{\eta_2+2\eta_3+\delta_{46}}+e^{\eta_1+\eta_2
		+\eta_3+\delta_{47}}\nonumber\\
	&&+e^{2(\eta_1+\eta_2)+\nu_{1}}+e^{2(\eta_1+\eta_3)+\nu_{2}}+e^{2(\eta_2+\eta_3)+\nu_{3}}+e^{2\eta_1+\eta_2+\eta_3+\nu_4}\nonumber\\
	&&+e^{\eta_1+2\eta_2+\eta_3+\nu_5}+e^{\eta_1+\eta_2+2\eta_3+\nu_6}+e^{2(\eta_1+\eta_2)+\eta_3+\nu_{11}}+e^{2(\eta_1+\eta_3)+\eta_2+\nu_{12}}\nonumber\\
	&&+e^{2(\eta_2+\eta_3)+\eta_1+\nu_{13}}+e^{2(\eta_1+\eta_2+\eta_3)+\nu_{21}}, \label{21b} 
\end{eqnarray}\ees
where $\eta_j=k_jx+\frac{k_j^3}{4}t+\eta_{j}^{(0)}$, $j=1,2,3,$ and the other constants are defined below:
\begin{eqnarray*}
&&e^{\del_j}= -\frac{2\mu}{k_j^2}, ~ e^{\Del_j}=\frac{\rho_j}{4k_j^4},~ \rho_j=(4\mu^2-k_j^2), ~j=1,2,3,~ e^{\chi_1}=-\frac{2\mu(k_1-k_2)^2}{k_1^2k_2^2},\nonumber\\
&&e^{\chi_2}=-\frac{2\mu(k_1-k_3)^2}{k_1^2k_3^2},~ e^{\chi_3}=-\frac{2\mu(k_2-k_3)^2}{k_2^2k_3^2},~ e^{\delta_{11}}=\frac{2\big[2\mu^2(k_1^2+k_2^2)-k_1^2k_2^2\big]}{k_1^2k_2^2(k_1+k_2)^2}, \nonumber\\
&&e^{\delta_{12}}=\frac{2\big[2\mu^2(k_1^2+k_3^2)-k_1^2k_3^2\big]}{k_1^2k_3^2(k_1+k_3)^2},~ e^{\delta_{13}}=\frac{2\big[2\mu^2(k_2^2+k_3^2)-k_2^2k_3^2\big]}{k_2^2k_3^2(k_2+k_3)^2},~e^{\chi_{11}}=\frac{A_{12}\rho_1}{4k_1^4},\nonumber\\
&&e^{\chi_{12}}=\frac{A_{12}\rho_2}{4k_2^4},~e^{\chi_{13}}=\frac{A_{13}\rho_1}{4k_1^4},~e^{\chi_{14}}=\frac{A_{13}\rho_3}{4k_3^4},~e^{\chi_{15}}=\frac{A_{23}\rho_2}{4k_2^4},~e^{\chi_{16}}=\frac{A_{23}\rho_3}{4k_3^4},\nonumber\\
&&e^{\chi_{17}}=\frac{2\lambda_1}{k_1^2k_2^2k_3^2(k_1+k_2)^2(k_1+k_3)^2(k_2+k_3)^2},~e^{\delta_{41}}=-\frac{\mu\rho_1A_{12}}{2k_1^4k_2^2},\nonumber\\
&&\lambda_1=\big(2\mu^2\big[k_1^6(k_2^2+k_3^2)+k_2^2k_3^2(k_2^2-k_3^2)^2-2k_1^4(k_2^4+k_3^4)+k_1^2(k_2^6+k_3^6)\big]\nonumber\\
&&\hspace{0.5cm}-k_1^2k_2^2k_3^2\big[k_1^4+k_2^4-k_2^2k_3^2+k_3^4-k_1^2(k_2^2+k_3^2)\big]\big), \nonumber\\
&&e^{\delta_{42}}=-\frac{\mu\rho_2A_{12}}{2k_1^2k_2^4},~e^{\delta_{43}}=-\frac{\mu\rho_1A_{13}}{2k_1^4k_3^2},~e^{\delta_{44}}=-\frac{\mu\rho_3A_{13}}{2k_1^2k_3^4},~e^{\delta_{45}}=-\frac{\mu\rho_2A_{23}}{2k_2^4k_3^2},\nonumber\end{eqnarray*}\begin{eqnarray*}
&&e^{\delta_{46}}=-\frac{\mu\rho_3A_{23}}{2k_2^2k_3^4},~e^{\delta_{47}}=\frac{4\mu\lambda_2}{k_1^2k_2^2k_3^2(k_1+k_2)^2(k_1+k_3)^2(k_2+k_3)^2},\\
&&e^{\chi_{41}}=-\frac{\mu\rho_1A_{12}A_{13}(k_2-k_3)^2}{2k_1^4k_2^2k_3^2},~e^{\chi_{42}}=-\frac{\mu\rho_2A_{12}A_{23}(k_1-k_3)^2}{2k_1^2k_2^4k_3^2},\\
&&\lambda_2=\big[-2\mu^2(k_1^4(k_2^2+k_3^2)+k_2^2k_3^2(k_2^2+k_3^2)+k_1^2(k_2^4-6k_2^2k_3^2+k_3^4))\\
&&\hspace{1.0cm}+(k_2^4k_3^4-k_1^2k_2^2k_3^2(k_2^2+k_3^2)+k_1^4(k_2^4-k_2^2k_2^2+k_3^4))\big],\\
&&e^{\chi_{43}}=-\frac{\mu\rho_3(k_1-k_2)^2A_{13}A_{23}}{2k_1^2k_2^2k_3^4},~e^{\nu_{1}}=\frac{\rho_1\rho_2A_{12}^2}{16k_1^4k_2^4},~e^{\nu_{2}}=\frac{\rho_1\rho_3A_{13}^2}{16k_1^4k_3^4},\\
&&e^{\nu_{3}}=\frac{\rho_2\rho_3A_{23}^2}{16k_2^4k_3^4},~e^{\nu_4}=\frac{\rho_1A_{12}A_{13}\big[2\mu^2(k_2^2+k_3^2)-k_2^2k_3^2\big]}{2k_1^4k_2^2k_3^2(k_2+k_3)^2},\nonumber\\
&&e^{\nu_5}=\frac{\rho_2A_{12}A_{23}\big[2\mu^2(k_1^2+k_3^2)-k_1^2k_3^2\big]}{2k_1^2k_2^4k_3^2(k_1+k_3)^2},~e^{\nu_6}=\frac{\rho_3A_{13}A_{23}\big[2\mu^2(k_1^2+k_2^2)-k_1^2k_2^2\big]}{2k_1^2k_2^2k_3^4(k_1+k_2)^2},\\
&&e^{\nu_{11}}=-\frac{\mu A_{12}^2A_{13}A_{23}\rho_1\rho_2}{8k_1^4k_2^4k_3^2}, ~e^{\nu_{12}}=-\frac{\mu A_{12}A_{13}^2A_{23}\rho_1\rho_3}{8k_1^4k_2^2k_3^4},~e^{\nu_{13}}=-\frac{\mu A_{12}A_{13}A_{23}^2\rho_2\rho_3}{8k_1^2k_2^4k_3^4},\\
&&e^{\gamma_{1}}=\frac{A_{12}^2A_{13}A_{23}\rho_1\rho_2}{16k_1^4k_2^4},~e^{\gamma_{2}}=\frac{A_{12}A_{13}^2A_{23}\rho_1\rho_3}{16k_1^4k_3^4},~e^{\gamma_{3}}=\frac{A_{12}A_{13}A_{23}^2\rho_2\rho_3}{16k_2^4k_3^4}, \end{eqnarray*}\begin{eqnarray*}
&&e^{\nu_{21}}=\frac{A_{12}^2A_{13}^2A_{23}^2\rho_1\rho_2\rho_3}{64k_1^4k_2^4k_3^4}, ~\text{and}~~A_{ij}=\frac{(k_i-k_j)^2}{(k_i+k_j)^2},~i>j,~i,j=1,2,3.
\end{eqnarray*}
The above three-soliton solution is characterized by six real constants $k_j$, and $\eta_j^{(0)}$, $j=1,2,3$, along with  the system parameter $\mu$. A detailed asymptotic analysis is given below to verify the underlying collision among the three solitons, as it is demonstrated in Fig. \ref{fg3}(a). 
\subsubsection{Collision among three solitons: Asymptotic analysis}
To analyze the collision among the three Gardner solitons in the asymptotic time limits $t\rightarrow \mp \infty$, a parametric choice, $k_1^2<k_2^2<k_3^2$, is considered in our asymptotic analysis. This choice corresponds to an overtaking collision among the solitons. By incorporating the asymptotic behaviour of the  wave variables,  $\eta_j=k_j(x+\frac{k_j^2}{4}t+\frac{\eta_{j}^{(0)}}{k_j})$, $j=1,2,3,$ in the three-soliton solution (\ref{21a} )-(\ref{21b}) one can deduce the asymptotic forms of each of the solitons. In the large time limits, $t\rightarrow\pm\infty$, the asymptotic behavior of the wave variables $\eta_j$'s are identified as (i) Soliton 1: $\eta_{1}\simeq 0$, $\eta_{2,3}\rightarrow\pm \infty$ as $t\rightarrow \mp\infty$, (ii) Soliton 2: $\eta_{2}\simeq 0$, $\eta_{1}\rightarrow\mp \infty$, $\eta_{3}\rightarrow\pm \infty$ as $t\rightarrow\mp\infty$, and (iii) Soliton 3: $\eta_{3}\simeq 0$, $\eta_{1,2}\rightarrow\mp \infty$, as $t\rightarrow\mp\infty$. By considering these results in the solution see Eq. (\ref{21a})-(\ref{21b}) we deduce the following asymptotic forms for each individual solitons. \\\\
{\bf (a) Before collision: $t\rightarrow -\infty$}\\
{\bf Soliton 1:} In this limit, the asymptotic form is deduced from the three-soliton solution (\ref{21a})-(\ref{21b}) for soliton 1 as below: 
\bea
v(x,t)=\frac{k_1^2}{-2\mu+\sqrt{4\mu^2-k_1^2}\cosh(\eta_1+\phi_1^-)},
\eea
where the phase term $\phi_1^{-}=\frac{1}{2}\log\frac{(k_1-k_2)^4(k_1-k_3)^4(4\mu^2-k_1^2)}{4k_1^4(k_1+k_2)^4(k_1+k_3)^4}$.
In the latter, superscript ($-$) represents before collision and subscript $1$ denotes soliton $1$.\\
{\bf Soliton 2:} The asymptotic form of soliton 2 is obtained as
\bea
v(x,t)=\frac{k_2^2}{-2\mu+\sqrt{4\mu^2-k_2^2}\cosh(\eta_2+\phi_2^-)},
\eea
where,  $\phi_2^{-}=\frac{1}{2}\log\frac{(k_2-k_3)^4(4\mu^2-k_2^2)}{4k_2^4(k_2+k_3)^4}$. Here, the subscript $2$ indicates soliton 2. \\ 
{\bf Soliton 3: }The following asymptotic form of soliton 3 is deduced from the solution (\ref{21a})-(\ref{21b}). This form is given by
\bea
v(x,t)=\frac{k_3^2}{-2\mu+\sqrt{4\mu^2-k_3^2}\cosh(\eta_3+\phi_3^-)}.
\eea
In the above, subscript $3$ in the phase shift 
$\phi_3^{-}=\frac{1}{2}\log\frac{4\mu^2-k_3^2}{4k_3^4}$ represents soliton 3. 
\\
{\bf (b) After collision: $t\rightarrow +\infty$}\\
{\bf Soliton 1:} Similarly, in this long time limit, the asymptotic forms of soliton 1 is obtained as
\bea
v(x,t)=\frac{k_1^2}{-2\mu+\sqrt{4\mu^2-k_1^2}\cosh(\eta_1+\phi_1^+)},
\eea where $\phi_1^{+}=\frac{1}{2}\log\frac{4\mu^2-k_1^2}{4k_1^4}$. In the latter, superscript $(+)$ denotes after collision. 
\\
{\bf  Soliton 2:} For soliton 2, the asymptotic expression  turns out to be
\bea
v(x,t)=\frac{k_2^2}{-2\mu+\sqrt{4\mu^2-k_2^2}\cosh(\eta_2+\phi_2^+)}.
\eea Here, $\phi_2^{+}=\frac{1}{2}\log\frac{(k_1-k_2)^4(4\mu^2-k_2^2)}{4k_2^4(k_1+k_2)^4}$. 
\\
{\bf Soliton 3:} The asymptotic expression for soliton 3 is deduced as 
\bea
v(x,t)=\frac{k_3^2}{-2\mu+\sqrt{4\mu^2-k_3^2}\cosh(\eta_3+\phi_3^+)}.
\eea 
Here, phase shift term $\phi_3^{+}$ is found to be  $\phi_3^{+}=\frac{1}{2}\log\frac{(k_1-k_3)^4(k_2-k_3)^4(4\mu^2-k_3^2)}{4k_3^4(k_1+k_3)^4(k_2+k_3)^4}$. \\

The asymptotic analysis clearly shows that the structures of the Gardner solitons remain invariant under collision. As it is evident from Fig. \ref{fg3}(a), the shapes of each of the solitons are preserved during the entire collision scenario thereby ensuring the shape preserving collision nature. However, each of the solitons acquires a phase shift. They  are given by
$\Delta \Phi_1=\frac{1}{2}\log\frac{(k_1+k_2)^4(k_1+k_3)^4}{(k_1-k_2)^4(k_1-k_3)^4}$, $\Delta\Phi_2=\frac{1}{2}\log\frac{(k_1-k_2)^4(k_2+k_3)^4}{(k_1+k_2)^4(k_2-k_3)^4}$, and
$\Delta \Phi_3=\frac{1}{2}\log\frac{(k_2-k_3)^4(k_1-k_3)^4}{(k_2+k_3)^4(k_1+k_3)^4}$.

\end{document}